\PassOptionsToPackage{table}{xcolor}

\documentclass[screen]{acmart}

\usepackage[nameinlink]{cleveref}  
\usepackage{dirtytalk}
\usepackage{enumitem}
\usepackage{longtable}
\usepackage{multirow}

\copyrightyear{2026}
\acmYear{2026}
\setcopyright{cc}
\setcctype{by}
\acmJournal{TKDD}
\acmVolume{To appear}
\acmDOI{10.1145/3798096}

\newcommand{\numpublishedpapers}{1125} 
\newcommand{\numpublishedpaperstext}{1,125}
\newcommand{\numexcludedpapers}{49}
\newcommand{\numcandidatepapers}{1076} 
\newcommand{\numcandidatepaperstext}{1,076}
\newcommand{\numselectedpapers}{204}   
\newcommand{\numselecteddatasets}{172} 

\renewcommand{\arraystretch}{1.25}

\begin{document}

\title{Open Datasets in Learning Analytics: Trends, Challenges, and Best PRACTICE}

\author{Valdemar Švábenský}
\orcid{0000-0001-8546-280X}
\affiliation{
    \department{Faculty of Informatics}
    \institution{Masaryk University}
    \city{Brno}
    \country{Czech Republic}
}
\email{valdemar@mail.muni.cz}

\author{Brendan Flanagan}
\orcid{0000-0001-7644-997X}
\affiliation{
    \institution{Kyoto University}
    \city{Kyoto}
    \country{Japan}
}
\email{bflanagan.academic@gmail.com}

\author{Erwin Daniel López Zapata}
\orcid{0000-0003-3793-9524}
\affiliation{
    \institution{Kyushu University}
    \city{Fukuoka}
    \country{Japan}
}
\email{edlopez96s6@gmail.com}

\author{Atsushi Shimada}
\orcid{0000-0002-3635-9336}
\affiliation{
    \institution{Kyushu University}
    \city{Fukuoka}
    \country{Japan}
}
\email{atsushi@ait.kyushu-u.ac.jp}

\begin{abstract}
\textbf{Background and context:}
Open datasets play a crucial role in three prominent research domains that intersect data science and education: learning analytics, educational data mining, and artificial intelligence in education. Researchers in these domains apply computational methods to analyze data from educational contexts, aiming to better understand and improve teaching and learning.\\
\textbf{Research scope and gap:}
Providing open datasets alongside research papers supports research reproducibility, fosters collaboration, and increases trust in research findings. It also provides individual benefits for authors, such as greater visibility, credibility, and citation potential. However, despite these advantages, the availability of open datasets and the associated practices within the learning analytics research communities, especially at their flagship conference venues, remain unclear.\\
\textbf{Goal and method:}
To address this gap, we conducted a systematic survey of publicly available datasets published alongside research papers in learning analytics domains. We manually examined \numpublishedpaperstext\ papers from three respected flagship conferences (LAK, EDM, and AIED) over the past five years (2020–2024). We discovered, categorized, and analyzed \numselecteddatasets\ unique datasets used in \numselectedpapers\ publications.\\
\textbf{Results and contributions:}
Our study presents the most comprehensive collection and analysis of open educational datasets to date, along with the most detailed categorization. Of the \numselecteddatasets\ datasets identified, 143 were not captured in any prior survey of open data in learning analytics. We provide insights into the datasets’ context, analytical methods, use, and other properties. Based on this survey, we summarize the current gaps in the field. Furthermore, we list practical recommendations, advice, and 8-item guidelines under the acronym PRACTICE with a checklist to help researchers publish their data. Lastly, we share our original dataset: an annotated inventory detailing the discovered datasets and the corresponding publications. We hope these findings will support further adoption of open data practices in learning analytics communities and beyond.
\end{abstract}



\received{26 February 2025}
\received[revised]{23 December 2025}
\received[accepted]{8 February 2026}

\maketitle

\section{Introduction}
\label{sec:intro}

Open science practices~\cite{Vicente2018, Kathawalla2021easing} are promoted by numerous respected organizations worldwide, such as UNESCO~\cite{unesco-openscience}, Center for Open Science~\cite{cos-openscience}, and European Commission~\cite{ec-openscience}. One of the key aspects of open science is the sharing of research data, which benefits various target groups~\cite{open-data-publish-springer, open-data-publish-elsevier, Science2023, Kathawalla2021easing}:
\begin{itemize}
    \item For \textit{individual researchers}, it brings greater recognition and visibility, leads to higher impact and more citations (estimated by +25\%~\cite{Colavizza2020_citation_advantage}), and increases their credibility and transparency.
    \item For \textit{research communities}, it supports reproducibility, enables new collaboration opportunities, and saves time and other resources in future studies, leading to faster progress.
    \item Finally, for \textit{society} in the broader sense, it enhances trust in research, informs policy, and simplifies access to research results for the industry and general public.
\end{itemize}

The popular guideline for publishing scientific data~\cite{wilkinson2016fair} (with more than 20,000 citations on Google Scholar a decade after its publication) states that the data should be FAIR~\cite{wilkinson2016fair, EU2018fair}:
\begin{itemize}
    \item \textit{Findable} -- have a persistent identifier, are indexed in a searchable resource, and are described with metadata.
    \item \textit{Accessible} -- are available and retrievable using standard means (by humans and machines).
    \item \textit{Interoperable} -- are syntactically parseable and semantically understandable, allowing to remix them~\cite{open-data-handbook}.
    \item \textit{Reusable} -- meet community standards and are released with an open license that allows wide usage.
\end{itemize}
Failing to meet the FAIR guidelines can negatively impact research activities and innovation. A~2018 report estimated that in the European Union alone, \say{not having FAIR and free open access research data [\ldots] costs the European economy at least €10.2 billion every year}~\cite{EU2018fair}.

Large-scale studies on open data practices revealed further insights. In 2018, a global survey of 7,719 researchers found that 76\% considered it important to make their data discoverable~\cite{Stuart2018}. However, the respondents identified numerous obstacles to this practice, namely: \textit{how to organize the data} (46\%), \textit{how to deal with licensing} (37\%), \textit{which repository to use} (33\%), \textit{lack of time to share data} (26\%), and \textit{costs of sharing data} (19\%). Another study~\cite{Tedersoo2021} demonstrated the inefficiency of explicitly requesting data. It takes a long time to receive a response, and ultimately, it is often declined -- the top three reasons being again \textit{lack of time} (29\%), \textit{data no longer available} (28\%), and \textit{privacy/legal concerns} (23\%).

A 2023 study~\cite{Science2023} with 6,091 responses revealed more concerns: 60\% of researchers felt they \say{don't receive sufficient recognition} for sharing their data, and almost 75\% \say{had never received support} with data publishing. The study also highlighted differences among subject areas, necessitating a domain-specific approach. Out of 13 areas (which included, for example, mathematics, engineering, and social sciences), researchers in \textit{computing}\footnote{Computing is a broad area that includes \say{computer science and related disciplines, such as computer engineering, information systems, information technology, software engineering, cybersecurity, and data science}~\cite{Guzdial2018}.} were among those who most strongly agreed that open data \say{should be common scholarly practice}. In addition, computing researchers were \say{significantly more likely to select paper citations as their primary motivation for sharing data}.

\subsection{Three Research Fields of Learning Analytics, Educational Data Mining, and AI in Education}
\label{subsec:intro-la}

At the intersection of computing and education lie three interrelated research disciplines: \textit{learning analytics}~\cite{handbook2017-la, romero2020}, \textit{educational data mining}~\cite{handbook2010-edm, romero2020}, and \textit{artificial intelligence (AI) in education}~\cite{Chen2020-aied, Chen2022_two_decades_AIED}. The goal of these interdisciplinary fields is to better understand and improve teaching and learning~\cite{romero2020, hundhausen2017}. To achieve this goal, the researchers leverage data from educational contexts and automatically analyze them using computational methods~\cite{Svabensky2022applications}, such as data mining, machine learning (ML), and natural language processing (NLP).

The research interest in these fields has been growing over the past years, as evidenced by the increasing number of publications for learning analytics~\cite{romero2020, Cerezo2024, Svabensky2026fifteen}, educational data mining~\cite{romero2020, Cerezo2024}, and AI in education~\cite{Chen2020-aied, Chen2022_two_decades_AIED}. All three fields have well-established communities -- \Cref{tab:conferences} lists the flagship conference for each field. These conferences are international, stringently peer-reviewed, and ICORE-ranked~\cite{core2023}. Moreover, LAK is the only conference in the top 20 Google Scholar citation ranks for the subcategory Educational Technology under Engineering \& Computer Science~\cite{google_scholar_LAK}. These metrics place LAK on par with the top journals in the field.

\begin{table}[t]
\caption{The most prominent conferences in each field. Data are bound to the time of writing this paper (submitted February 2025). We note that while LAK and EDM have been held annually since their inception, AIED has been held biannually since 1983, then switching to an annual mode since 2017~\cite{self2016birth}.}
\label{tab:conferences}

\renewcommand{\arraystretch}{1.8}
\setlength{\tabcolsep}{4.4pt}
\begin{tabular}{p{5.5cm}p{4.6cm}p{1.2cm}p{0.7cm}p{1.8cm}}
\toprule
\textbf{Conference name (with abbreviation) and URL} & \textbf{Publisher} & \textbf{ICORE rank~\cite{core2023}} & \textbf{First year} & \textbf{Anniversary in 2025} \\ 
\midrule

International Learning Analytics and Knowledge Conference (\textbf{LAK}) {\small\url{https://www.solaresearch.org/events/lak}} & Society for Learning Analytics Research (SoLAR) and Association for Computing Machinery (ACM) & A & 2011 & 15\textsuperscript{th}\\

Educational Data Mining (\textbf{EDM}) \hspace{1cm} {\small\url{https://educationaldatamining.org/conferences}} & The International Educational Data Mining Society (IEDMS) & B & 2008 & 18\textsuperscript{th} \\

Artificial Intelligence in Education (\textbf{AIED}) {\small\url{https://iaied.org/conferences}} & The International AI in Education Society and Springer & A & 1983 & 26\textsuperscript{th}~\cite{self2016birth} \\

\bottomrule
\end{tabular}

\end{table}


For readers interested in details, \citet{romero2020} published a comprehensive survey of the state of the art in learning analytics and educational data mining, explaining the fine-grained differences between the two and outlining their relationship to AI in education. However, these details are not particularly relevant for the scope of our paper, which targets all three disciplines. Moreover, \citet{Dormezil2019} noted that, from one perspective, educational data mining can be considered a subset of learning analytics. Therefore, for simplicity, this paper will use the umbrella term \textit{Learning Analytics} (LA) to refer to all three disciplines combined. Only when referring to a specific conference will we use the respective acronyms LAK, EDM, and AIED (see \Cref{tab:conferences}).

\subsection{Goals and Target Audience of This Paper}

This paper uses the simplified working definition of open data listed below. It combines the FAIR guidelines~\cite{wilkinson2016fair, EU2018fair} with one of the original open data definitions~\cite{murray2008open}, as well as with the definition by the Open Data Handbook~\cite{open-data-handbook}.
\begin{quote}
\textit{Open data are data that (1) are easily available in a standard format, and (2) can be used by anyone for any (non-commercial) research purpose --- subject at most to the requirement to attribute and/or share-alike.}
\end{quote}

Within the scope of this definition, our goal is to conduct a practically oriented survey of open data for LA research. We examine all \numpublishedpaperstext\ papers published at LAK, EDM, and AIED from 2020 to 2024 (see \Cref{subsec:method-scope} for why we set this scope). Reflecting our paper title, we aim to answer the following research questions (RQ):
\begin{enumerate}
    \item[(RQ1)] \textbf{Trends:} \textit{Which open datasets are available in state-of-the-art LA research, and what are their characteristics?}
    \item[(RQ2)] \textbf{Challenges:} \textit{Which contexts does the LA community lack data for, representing research gaps and open problems?}
    \item[(RQ3)] \textbf{Best Practice:} \textit{How can we guide and support LA researchers to adopt more open data practices?}
\end{enumerate}

The intended audience of this paper is researchers and other stakeholders in LA who want to:
\begin{itemize}
    \item Adopt existing datasets as-is to avoid the \say{cold start} problem (that is, not having enough research data at the beginning), enabling them to kickstart their new research faster.
    \item Adapt existing datasets to suit their specific needs, for example, as a test set for evaluating their ML models.
    \item Study the factors that affect or limit the generalizability of research findings across datasets.
    \item Validate the previously published research through replication or reproduction.
    \item Seek inspiration for unanswered research problems or areas that are insufficiently covered by open data.
    \item Look for practical, easy-to-use guidelines to help them engage in open data practices.
\end{itemize}

\subsection{Contributions of This Paper}

Our paper brings the following contributions to the research community:
\begin{enumerate}
    \item \textit{An organized resource with pointers to and descriptions of open LA datasets} that others can use (\Cref{sec:appendix-datasets}).
    \item \textit{Comparison with previous surveys}, providing references to datasets that our survey did not cover (\Cref{subsec:related-work-data}).
    \item \textit{Summary of trends} regarding the dataset properties, educational contexts, and analytical methods (\Cref{subsec:results-summary}).
    \item \textit{Critical reflection of the state of the art} in LA research and related disciplines, leading to the \textit{identification of research gaps and open challenges} to provide opportunities for novel contributions (\Cref{subsec:implications-future-work}).
    \item \textit{Actionable recommendations} and implications for LA researchers regarding datasets (\Cref{subsec:implications-recommendations} and \Cref{sec:appendix-checklist}).
    \item \textit{Public materials} including the dataset inventory, analytical code, and structured citation files (\Cref{subsec:implications-materials}).
\end{enumerate}

\section{Related Literature Reviews and Surveys in Learning Analytics}

We examined prior work focusing on two aspects: data (\Cref{subsec:related-work-data}) and open science in general (\Cref{subsec:related-work-open-science}). As a reminder, we use the term LA to refer jointly to learning analytics, educational data mining, and/or AI in education.

\subsection{Usage of Data in LA}
\label{subsec:related-work-data}

\subsubsection{Broad Survey and Overview Papers}

In 2020, \citet{romero2020} thoroughly surveyed the field of LA, pointing out that collecting and preprocessing data for LA research is a \say{hard and very time-consuming task}. They observed that few public datasets are available, and that they are limited to a narrow range of educational environments (mostly e-learning systems). They listed URLs to 13 public datasets that can be used by LA researchers. Of these URLs, four are invalid at the time of writing this paper, underscoring the importance of using stable data repositories. 

Data collection in LA is particularly complex since it often involves a multi-layered approach to capture a complete picture of how learning is occurring, the external factors that influence it, and its outcomes. This issue was examined by \citet{Fischer2020}, who reviewed how the field of LA employed big data, dividing them into three types:
\begin{itemize}
    \item \textit{Microlevel}: fine-grained interaction data between learners and the learning environment, e.g., clickstream data. They were used, for example, to understand knowledge or personalize learning.
    \item \textit{Mesolevel}: student writing artifacts, e.g., text data from online discussions. They were used, for example, to understand cognitive, social, and behavioral processes.
    \item \textit{Macrolevel}: institutional data, e.g., demographics and course enrollment data. They were used, for example, in early-warning systems, for course guidance, and analytics for administrators and decision-makers.
\end{itemize}
This review by \citet{Fischer2020} posits that accessing educational big data is a major challenge. Moreover, educational software companies, which possess large datasets, \say{have no interest in making their data available publicly}~\cite{Fischer2020} in order to retain their commercial advantage.

The costs of data collection and preprocessing, combined with the lack of open LA data, create a digital divide that widens inequality within the research community. As a step towards addressing this issue, \citet{romero2020} proposed developing an open repository specific to LA data, such as the PSLC DataShop~\cite{handbook2010-edm}. They conclude that \say{educators and institutions should develop a data-driven culture of using data for [\ldots] improving instruction}.

\subsubsection{Dataset Survey Papers and Dataset Repositories}

\citet{Mihaescu2021review} identified 44 LA datasets, superseding the data aspect of the review by \citet{romero2020}. Almost all of the datasets were published between 2010 and 2020 (with two exceptions from 1988 and 1997). Since our review focuses on papers from 2020 to 2024, we provide a follow-up to this previous survey. \citet{Mihaescu2021review} searched for papers and datasets using Google Scholar and Google Dataset Search~\cite{google_dataset_search}, and categorized the dataset sources into three types~\cite{Mihaescu2021review}:
\begin{itemize}
    \item General-purpose repositories:
    (a) UCI Machine Learning Repository~\cite{UCI_ML_repo}, 
    (b) Mendeley Data Repository~\cite{Mendeley_data_repo}, 
    (c) Harvard Dataverse~\cite{Harvard_data_repo}, 
    (d) PSLC DataShop~\cite{datashop, datashop_website}, 
    and, as we would like to additionally include in this category, (e) LearnSphere~\cite{Learnsphere_website}.
    \item Datasets from LA competitions, which are often hosted on Kaggle~\cite{Kaggle_website}. 
    \item Standalone datasets from other sources.
\end{itemize}

Next, based on a Google Scholar keyword search, we identified three other surveys that describe datasets in LA, all of which substantially overlap with each other and with prior work (see \Cref{tab:related-surveys}). Listed chronologically, the first is \citet{bigdata2021edudata}: a GitHub repository with an interface for downloading 18 LA datasets (15 of which are new compared to prior surveys). Second, \citet{lin2024surveydeeplearning} listed 10 datasets (1 new compared to prior work). Third, \citet{Xiong2024survey} listed 12 datasets (2 of which are new compared to prior work). Other than that, to the best of our knowledge, our paper is the largest and most comprehensive survey of LA datasets to date, with \numselecteddatasets\ datasets in total.

To highlight our contribution, \Cref{tab:related-surveys} compares our work with other surveys. Combined together, all previous surveys reported 62 unique datasets. Of these, 29 were also identified in our paper, and 33 were not. However, we identified 143 additional datasets missing from previous surveys.

\begin{table}[t]

\caption{Comparison of our work to the related surveys reviewed in \Cref{subsec:related-work-data}, sorted by publication date. We do not list the surveys by Haim et al.~\cite{Haim2023lak, Haim2023edm, Haim2023aied} because they examined a subset of our candidate papers.}
\label{tab:related-surveys}

\rowcolors{2}{gray!15}{white}
\begin{tabular}{lp{1.5cm}p{1.25cm}p{3.25cm}p{3.7cm}}
\toprule
\textbf{Survey of public datasets} & \textbf{Publication date} & \textbf{Datasets reported} & \textbf{Overlap with datasets across all prior work} & \textbf{Overlap with datasets in our survey (see \Cref{tab:related-surveys-overlap})} \\ 
\midrule

\citet{romero2020}                & 2020, Jan & 12 & --- (the first survey) & 8 \\

\citet{Mihaescu2021review}        & 2021, Feb & 44 & 12 (27\%), superset of~\cite{romero2020} & 17 \\

\citet{bigdata2021edudata}        & 2021, Nov & 18 & 3 (17\%) & 14 \\

\citet{lin2024surveydeeplearning} & 2024, Jun & 10 & 9 (90\%) & 9 \\

\citet{Xiong2024survey}           & 2024, Jul & 12 & 10 (83\%) & 9 \\

Our paper                         & 2025--2026 & \textbf{\numselecteddatasets} & 29 (17\%) & --- \\

\bottomrule
\end{tabular}

\bigskip
\bigskip

\caption{Explanation of the column ``Overlap with datasets in our survey'' from \Cref{tab:related-surveys}. The numbers refer to dataset IDs in \Cref{tab:datasets}.}
\label{tab:related-surveys-overlap}

\renewcommand{\arraystretch}{1.5}
\setlength{\tabcolsep}{3.5pt}
\tiny
\begin{tabular}{lccccccccccccccccccccccccccccc}
\toprule
\rowcolor{white}
\textbf{Survey of public datasets} & 2 & 3 & 5 & 6 & 18 & 19 & 20 & 22 & 23 & 30 & 41 & 42 & 50 & 68 & 75 & 83 & 86 & 103 & 105 & 109 & 127 & 134 & 140 & 149 & 161 & 168 & 169 & 170 & 171 \\
\midrule

\citet{romero2020} &  &  &  &  &  &  &  & \checkmark & \checkmark & \checkmark &  &  &  &  & \checkmark &  & \checkmark &  &  & \checkmark &  &  &  &  & \checkmark &  & \checkmark &  &  \\

\citet{Mihaescu2021review} & \checkmark & \checkmark & \checkmark & \checkmark &  &  &  & \checkmark & \checkmark & \checkmark & \checkmark & \checkmark & \checkmark &  & \checkmark &  & \checkmark &  &  & \checkmark & \checkmark &  &  & \checkmark & \checkmark &  & \checkmark &  &  \\

\citet{bigdata2021edudata} &  &  &  &  & \checkmark & \checkmark & \checkmark & \checkmark &  &  &  &  & \checkmark & \checkmark &  & \checkmark & \checkmark & \checkmark & \checkmark &  &  & \checkmark & \checkmark &  &  & \checkmark &  &  & \checkmark \\

\citet{lin2024surveydeeplearning} &  &  &  &  &  &  &  & \checkmark &  & \checkmark &  &  & \checkmark &  &  & \checkmark & \checkmark &  &  & \checkmark &  &  &  &  &  &  & \checkmark & \checkmark & \checkmark \\

\citet{Xiong2024survey} &  &  &  &  &  &  &  & \checkmark &  &  &  &  & \checkmark & \checkmark &  & \checkmark & \checkmark & \checkmark & \checkmark &  &  &  &  &  &  &  &  & \checkmark & \checkmark \\

\bottomrule
\end{tabular}

\end{table}

\subsubsection{Narrowly Focused, Area-Specific Survey Papers}

The rest of this subsection lists LA-related surveys that did not search for datasets directly but rather for related aspects, mainly approaches to dataset analytics.

Like our paper, \citet{Zheng2022} reviewed the LAK, EDM, and AIED conferences (additionally including a fourth conference, Learning at Scale). Among 66 core authors across the communities, more than 70\% hold doctoral degrees in computer science or engineering, highlighting a strong connection to computing. Moreover, while 90\% of core authors in AIED and EDM worked at institutions in the USA, 80\% of core authors in LAK worked in various countries outside the USA, demonstrating interest in LA worldwide. However, very few institutions in the Global South were represented.

\citet{Prenkaj2020} reviewed ML approaches for predicting student dropout in online education (such as in MOOCs, e-courses, and online degree programs). They listed several datasets used in the literature; however, only two were publicly available. The survey argued that due to the lack of available datasets, it is difficult to evaluate and compare (or otherwise benchmark) different ML models.

Next, \citet{sghir2023recent} reviewed 74 papers published over a decade of LA research on predicting student outcomes, such as enrollment and performance. Of the 74 papers, the review mentions only three that used public datasets (all three used the OULAD dataset -- ID 109 in our \Cref{tab:datasets}). The review asserts that sharing \say{more public datasets will surely accelerate the research on [predictive] LA}.

Lastly, \citet{Svabensky2022applications} reviewed 35 papers that had applied LA in cybersecurity education. They observed that the research data are highly heterogeneous, comprising nine distinct data types. These included program code, application logs, student communication data, and others. However, none of the datasets were made publicly available.

To summarize, several related publications argue for the importance of open datasets in LA and warn that they are lacking. However, existing surveys of open LA datasets are limited, either because they are outdated or because they list only a few datasets. To the best of our knowledge, our paper presents the most comprehensive review of LA datasets to date (finalized in February 2025).

\subsection{Open Science and Reproducibility of LA Research}
\label{subsec:related-work-open-science}

The Association for Computing Machinery (ACM) defines three key terms regarding the stable attainment of results from a research study~\cite{acm-repr}:
\begin{itemize}
    \item \textit{Repeatability} -- a research team can obtain their own result again, under the same conditions.
    \begin{itemize}
        \item \say{Same team, same experimental setup}
    \end{itemize}
    \item \textit{Reproducibility} -- a different team (other than the original authors) can independently obtain the original result by using data and artifacts from the original author team.
    \begin{itemize}
        \item \say{Different team, same experimental setup}
    \end{itemize}
    \item \textit{Replicability} -- a different team can obtain the result by using different data and artifacts, indicating that the phenomenon being researched is generalizable and manifests also under different conditions.
    \begin{itemize}
        \item \say{Different team, different experimental setup}
    \end{itemize}
\end{itemize}

From an open-science perspective, \textit{reproducibility} is especially important, as it enhances confidence in research results and enables broader applicability. However, a survey of 1,576 scientists across disciplines found that 90\% believe there is a reproducibility crisis~\cite{baker2016-repr}. One of the reasons, cited in almost 80\% of cases as \say{always/often/sometimes contributing} to the problem, is that \textit{raw data are not available from the original research team}~\cite{baker2016-repr}. To improve reproducibility practices in computing, a checklist for general ML research is available~\cite{Pineau2021}. The rest of this section reviews reproducibility specifically in the LA field, where it is also a challenge~\cite{Kitto2023}.

\subsubsection{Open Science at LAK, EDM, and AIED}

\citet{Motz2022} investigated open science practices in LAK 2020 and 2021 full and short papers. Of the 151 included articles, 11 (7.5\%) published their data. The data collection location was examined: 3 papers were from North America, 1 from Europe, and the rest were either multinational or unspecified.

In a related work, \citet{Haim2023lak} investigated the reproducibility of LAK 2021 and 2022 conference full and short papers. Only 5\% of papers made their raw dataset available, and in another 2\%, the paper explicitly stated that the data can be requested. Ultimately, no paper was reproducible within the time constraints set by \citet{Haim2023lak}, but they \say{estimated that the 2\% of papers that contain both the raw dataset and source [code] were likely to be reproducible}.

Next, \citet{Haim2023edm} researched the reproducibility of EDM 2021 and 2022 conference papers (full, short, and poster papers). The results showed that 15\% of papers used a dataset that was either already public or made openly available, and for another 5\%, the authors stated that the data could be requested. Ultimately, \citet{Haim2023edm} could attempt reproduction for 6\% of papers, but only a few yielded results similar to the original.

Last but not least, \citet{Haim2023aied} continued to explore the reproducibility of AIED 2021 and 2022 full, short, and poster papers. 13\% of papers had openly available data, and in 4\% of cases, the datasets were available on request. Overall, 7\% of papers were potentially reproducible, though none of the efforts were ultimately successful.

To summarize, open science practices are of interest at LAK, EDM, and AIED conferences. The topic of open science has been researched from a certain perspective in the years 2020--2022, but the latest results, with a deep focus on data, are currently missing. Our paper aims to fill this gap in the literature.

\subsubsection{Cautionary Viewpoints for LA Research}
\label{subsubsec:related-work-cautionary}

\citet{Baker2024open} pointed out that although open science is a worthwhile endeavor, publishing LA data can be difficult due to legal and institutional requirements, as well as issues regarding privacy, data security, and ethics. To make a dataset public, researchers need to~\cite{Baker2024open}:
\begin{itemize}
    \item[(a)] \textit{deidentify it} and 
    \item[(b)] \textit{obtain permission to share it}.
\end{itemize}

Deidentification~\cite{Campbell2023deidentifying, Guo2025} is challenging if the data includes photos, videos, audio recordings, or written communication of people~\cite{bosch2020hello}, but even in less obvious cases, such as with keystroke data~\cite{Leinonen2017}. To ensure data anonymization, a combined effort of several humans, aided by automated tools such as Large Language Models (LLMs)~\cite{Singhal2024deidentifying}, is necessary. However, even after deidentification, there remains the risk of reidentification, such as by combining the data with other datasets that contain personal information. To mitigate this risk, infrastructures such as MORF~\cite{Hutt2022} enable researchers to analyze data without direct access to them. Another possible mitigation is using synthetic or simulated data, which have recently been explored in the LA field~\cite{flanagan2022fine, Zhan2023synthetic, Kaser2024simulated}.

Obtaining permission is problematic in certain countries due to local legislation (see the discussion of \say{The right to be forgotten}~\cite{hutt2023right}), or at certain institutions -- even more so if the data come from a protected group, such as children. Again, infrastructures such as MORF~\cite{Hutt2022} partially mitigate this problem.

\section{Method of Conducting This Literature Survey}
\label{sec:method}

\begin{figure}[t]
\centering
\includegraphics[width=0.75\linewidth]{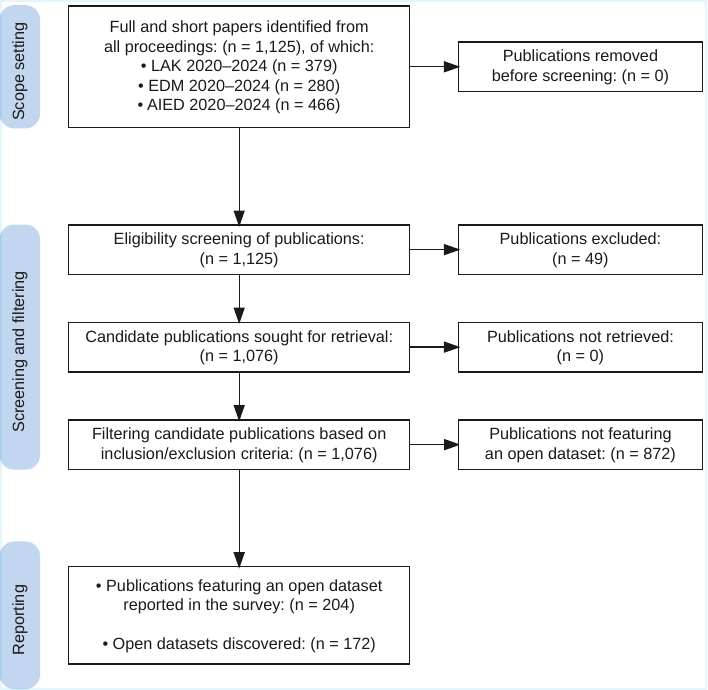}
\caption{PRISMA flow diagram. Generated using the tool by \citet{PRISMAtool2022}.}
\Description{Diagram showing the progression of selected papers.}
\label{fig:prisma}
\end{figure}

To reiterate, our goal is to conduct a practically-oriented survey of LA datasets. \Cref{fig:prisma} summarizes an overall PRISMA (Preferred Reporting Items for Systematic reviews and Meta-Analyses) flow diagram~\cite{PRISMA2021} for our literature survey. 

\subsection{Setting the Scope and Information Sources to Define the Initial Set of \numpublishedpaperstext\ Papers}
\label{subsec:method-scope}

Unlike most surveys, which examine a subset of publications yielded by a keyword search in a citation database, we aimed to capture an entire snapshot of the field, without risking that an incomplete keyword search would miss certain papers. Therefore, we focus on \textit{all} papers within the last 5 years in the flagship LA venues listed in \Cref{tab:conferences}.

We chose the LAK, EDM, and AIED conferences for the reasons explained in \Cref{subsec:intro-la} (to summarize: they are well-known, respected, and high-ranked). In addition, we selected conferences instead of journals because they capture the latest developments in the field. Given the longer publication turnaround in journals, it might take considerable time before datasets relevant to today's research start appearing there. In addition, conferences include more emerging research and topics outside the mainstream~\cite{Svabensky2026fifteen}, thereby more accurately reflecting the state of the art.

For each conference, we examined all full and short papers published between 2020 and 2024 (included). In total, \numpublishedpaperstext\ papers had been published at these conferences in this period. \Cref{tab:all-published-papers} shows a breakdown of the paper counts.

\begin{table*}
\caption{Distribution of all the \numpublishedpaperstext\ published papers (full | short) across the three target conferences throughout the years 2020--2024. In two cases, the text of the proceedings preface listed a slightly incorrect total paper count. Therefore, the numbers below are actual counts of all examined papers. (Each year, LAK published a total of 62--95 papers, EDM 40--66 papers, and AIED 76--116 papers.)}
\label{tab:all-published-papers}

\setlength\tabcolsep{4pt}

\begin{tabular}{l p{2mm} r|r p{2mm} r|r p{2mm} r|r p{2mm} r|r p{2mm} r|r p{2mm} r|r p{2mm} r}
\toprule
Full | Short & & \multicolumn{2}{c}{2020} & & \multicolumn{2}{c}{2021} & & \multicolumn{2}{c}{2022} & & \multicolumn{2}{c}{2023} & & \multicolumn{2}{c}{2024} & & \multicolumn{2}{c}{\textit{Subtotal}} & & \textbf{Total} \\
\midrule
LAK & &
50 & 30 & & 
42 & 29 & & 
39 & 23 & & 
49 & 22 & & 
66 & 29 & & 
\textit{246} & \textit{133} & & \textbf{379} \\
EDM & &
30 & 29 & & 
22 & 44 & & 
26 & 29 & & 
18 & 22 & & 
21 & 39 & & 
\textit{117} & \textit{163} & & \textbf{280} \\
AIED & &
49 & 66 & & 
40 & 76 & & 
40 & 40 & & 
53 & 26 & & 
49 & 27 & & 
\textit{231} & \textit{235} & & \textbf{466} \\
\textit{Subtotal} & &
\textit{129} & \textit{125} & & 
\textit{104} & \textit{149} & & 
\textit{105} & \textit{92} & & 
\textit{120} & \textit{70} & & 
\textit{136} & \textit{95} & & 
\textit{594} & \textit{531} & & \textbf{\numpublishedpapers} \\
\textbf{Total} & &
\multicolumn{2}{c}{\textbf{254}} & & 
\multicolumn{2}{c}{\textbf{253}} & & 
\multicolumn{2}{c}{\textbf{197}} & & 
\multicolumn{2}{c}{\textbf{190}} & & 
\multicolumn{2}{c}{\textbf{231}} & & 
\multicolumn{2}{c}{\textbf{\numpublishedpapers}} & \\
\bottomrule
\end{tabular}

\end{table*}

\subsection{Overall Workflow}

Data collection began on 20 May 2024 and ended on 17 October 2024. The first and second authors of this survey, who have an extensive expertise (Ph.D. in computer science) and publication track record in LA, divided all \numpublishedpaperstext\ papers between them. One co-author focused on LAK and AIED, while the other handled all EDM papers. In most cases, it was clear whether a dataset was present. Rarely, if either co-author encountered an unclear case, they discussed it with the team to reach a shared decision. Therefore, each paper was manually examined by one or more co-authors.

We attempted to complement and double-check our findings by AI-powered analysis of the papers' PDFs with Google's NotebookLM, but the outputs were inaccurate, so we resorted to a fully manual inspection. This corroborates \citet{Gehrmann2024_LLM_surveys}, who asserted that \say{while LLM-based tools can offer a useful initial overview of an unfamiliar topic, they are less effective for in-depth literature reviews}. Although LLM tools appear to be suitable for literature reviews, this use case \say{is still in its infancy}~\cite{Antu2023_LLM_surveys} and introduces challenges.

\subsection{Eligibility Screening of the Initial Set of \numpublishedpaperstext\ Papers}
\label{subsec:methods-eligibility}

As the initial step, the two investigators excluded all papers that did not exploit any dataset. These included position/opinion papers, proposals, and system design papers that lacked empirical evaluation. In addition, we excluded literature review/survey papers, since we wanted to focus on primary educational datasets. Overall, we excluded \numexcludedpapers\ papers (29 from LAK, 4 from EDM, and 13 from AIED). Since we excluded only 4.4\% out of the total \numpublishedpaperstext\ published papers, this shows that the vast majority of published papers were eligible to provide or use an open dataset. This left us with \numcandidatepaperstext\ candidate papers (see \Cref{tab:all-candidate-papers}) for the next round of review to explore.

\begin{table*}
\caption{Distribution of all the \numcandidatepaperstext\ candidate (eligible) papers. Same as in \Cref{tab:all-published-papers}, the papers are divided into full | short across the three target conferences throughout the years 2020--2024. In addition, the row and column \textbf{Total} include a percentage of the candidate papers compared to the corresponding cell in \Cref{tab:all-published-papers}.}
\label{tab:all-candidate-papers}

\setlength\tabcolsep{4pt}

\begin{tabular}{l p{2mm} r|r p{2mm} r|r p{2mm} r|r p{2mm} r|r p{2mm} r|r p{2mm} r|r p{2mm} r}
\toprule
Full | Short & & \multicolumn{2}{c}{2020} & & \multicolumn{2}{c}{2021} & & \multicolumn{2}{c}{2022} & & \multicolumn{2}{c}{2023} & & \multicolumn{2}{c}{2024} & & \multicolumn{2}{c}{\textit{Subtotal}} & & \textbf{Total} \\
\midrule
LAK & &
47 & 27 & & 
41 & 26 & & 
34 & 21 & & 
45 & 20 & & 
63 & 26 & & 
\textit{230} & \textit{120} & & \textbf{350} (92.3\%) \\
EDM & &
30 & 28 & & 
22 & 44 & & 
26 & 29 & & 
17 & 21 & & 
21 & 38 & & 
\textit{116} & \textit{160} & & \textbf{276} (98.6\%) \\
AIED & &
48 & 64 & & 
39 & 70 & & 
38 & 40 & & 
52 & 26 & & 
48 & 25 & & 
\textit{225} & \textit{225} & & \textbf{450} (96.6\%) \\
\textit{Subtotal} & &
\textit{125} & \textit{119} & & 
\textit{102} & \textit{140} & & 
\textit{98} & \textit{90} & & 
\textit{114} & \textit{67} & & 
\textit{132} & \textit{89} & & 
\textit{571} & \textit{505} & & \textbf{\numcandidatepapers} (95.6\%) \\
\textbf{Total} & &
\multicolumn{3}{l}{\textbf{244} (96.1\%)} & 
\multicolumn{3}{l}{\textbf{242} (95.7\%)} & 
\multicolumn{3}{l}{\textbf{188} (95.4\%)} & 
\multicolumn{3}{l}{\textbf{181} (95.3\%)} & 
\multicolumn{3}{l}{\textbf{221} (95.7\%)} & 
\multicolumn{3}{l}{\textbf{\numcandidatepapers} (95.6\%)} \\
\bottomrule
\end{tabular}

\end{table*}

\subsection{Examining the Eligible Set of \numcandidatepaperstext\ Candidate Papers}

The two investigators looked for any mention that the dataset associated with a paper may be available. We counted either providing a new, original dataset or exploiting an existing open dataset. Most frequently, the dataset was indicated by \textit{hyperlinks} in the text or \textit{citations} pointing to the data source: either a data storage (such as GitHub) or a previous paper. In the latter case, we recursively followed backward references if a current paper did not contain the dataset link directly.

Most often, the dataset reference appeared in the Methods section. Other locations included the Abstract, Introduction, Conclusion, or Appendix section, sometimes under a subheading titled, for example, \textit{dataset}, \textit{data set}, \textit{data collection}, \textit{supplemental/supplementary material(s)}, or \textit{availability of data/materials}.

Since papers and datasets have an \textit{m:n} correspondence -- that is, one paper can use multiple datasets, and one dataset can be used in multiple papers -- we distinguish between \textit{candidate papers} vs. \textit{candidate datasets}, as well as \textit{selected papers} vs. \textit{selected datasets}.

\subsection{Filtering the Candidate Papers and Datasets Based on Inclusion/Exclusion Criteria}
\label{subsec:methods-criteria}

After discovering a candidate dataset, we included it only if it satisfied all three of the following selection criteria:
\begin{enumerate}
    \item \textit{The dataset is indicated as being available/accessible (for example, the paper contains a functional hyperlink, a reference, a supplementary attachment in the proceedings, or a sentence such as \say{we will provide the data on request}).}
    \begin{itemize}
        \item Exclusion example 1: A paper only described the dataset in the text, but did not provide any link or a reference to it, and did not even indicate the possibility that it might be available on request.
        \item Exclusion example 2: A paper said \say{we used the same dataset as in [link/reference to a previous paper]}, but the referenced source again only textually described the dataset, that is, the data were again unavailable.
        \item Exclusion example 3: A paper described that the data are provided in a linked repository, but the repository was empty, and it seemed that it had never been filled (for example, no Git commit history).
        \item Exclusion example 4: A paper described that the data are provided in a linked storage, but the hyperlink was no longer functional (error 404), and its valid equivalent could not be found.\\
    \end{itemize}

    \item \textit{The dataset must be accessible electronically. The following ways to obtain it are possible:%
    \footnote{We based these options on Springer Nature guidelines~\cite{open-data-licenses-nature}, which distinguish: (a) \say{login-free https access}, (b) \say{basic login} with \say{immediate access [\dots] without manual checks}, or (c) \say{formal application processes} (which are discouraged by the guidelines of~\citet{open-data-licenses-nature}).}
    (a) direct download, (b)~download after registering a free-of-charge account to a hosting service and logging in, or (c) on request -- for anyone in the broader research community, for an unlimited time, and within 30 days of making the request.}
    \begin{itemize}
        \item Exclusion example 1: The dataset is indicated as available only to the members of a certain research group, university, or country. (This can be because of data privacy agreements.)
        \item Exclusion example 2: The licensing agreement to obtain a dataset mandates that the recipient must destroy their copy of the data within 30 days. (This may not allow others sufficient time for the analysis.)
        \item Exclusion example 3: Highly specialized technical skills and/or unreasonably time-consuming effort are required, such as deploying a codebase and running it to obtain the data. (This limits broader, direct use.)
        \item Exclusion example 4: After requesting a dataset via email or other official means, the request was ignored or took more than 30 days to receive a response.\\
    \end{itemize}
    
    \item \textit{The dataset was exploited to yield (at least partially) the results reported in the paper.}
    \begin{itemize}
        \item Exclusion example 1: A paper published a spreadsheet with the final results of the performed statistical tests, but the source dataset used as input for these computations to obtain the results is not available.
        \item Exclusion example 2: A paper published results based on numerous data points, but only one or a very few illustrative example data points are available.
        \item Exclusion example 3: A paper published code or a software tool, along with a \say{dummy} example dataset that enables running the code, but the paper analyzed a different (or much larger) dataset to report its results, and this dataset is not available.
    \end{itemize}
\end{enumerate}

Each paper was given at least 5 minutes (for full papers) or 3 minutes (for short papers) for a quick skim reading. After that, if it was obvious that some of the inclusion criteria were violated, we moved on to the next paper. If it was not clear within the time frame, or if additional sources (websites, backward references) needed to be inspected, the paper was allocated additional time. This way, we aimed to minimize the risk of false negatives (that is, accidentally excluding a paper that shared a dataset when that information was not immediately obvious).

For each dataset we requested, we used the same email template, addressing all authors of the associated paper. All emails were sent from the first author's institutional email address (that is, the official university account). If we did not receive a response within 15 days, we sent a reminder. If we did not receive a response within an additional 15 days, we excluded the dataset per our criterion (2c) above. Both email templates we used (the initial request and the reminder) are available in \Cref{subsec:implications-materials}.

Lastly, after completing data collection for the 2021 and 2022 papers, we also compared our findings with those of Haim et al. for LAK~\cite{Haim2023lak}, EDM~\cite{Haim2023edm}, and AIED~\cite{Haim2023aied}. We mostly identified the same datasets, or even additional ones, as those reported in these papers. However, on a few occasions, we were unable to find the dataset that Haim et al.~\cite{Haim2023lak, Haim2023edm, Haim2023aied} listed as being publicly available. Most likely, this was because several years had passed since the previous surveys, and the availability of the datasets had changed in the meantime.

\subsection{Extracting and Recording the Information About the Selected Papers and Datasets}

If a candidate dataset passed the inclusion criteria, it was recorded in a custom Google Sheets spreadsheet along with the associated paper. The spreadsheet structure was iteratively defined by the 1st author and updated based on the discussions with the 2nd author. Then, it was reviewed and updated by the 3rd and 4th authors, who did not participate in the initial data collection, to ensure independence. Finally, the 3rd author thoroughly validated, double-checked, and corrected the information in the spreadsheet, revisiting the source papers and datasets as needed. The final output is available alongside our supplementary materials -- see \Cref{subsec:implications-materials}.

\section{Survey Findings -- Available LA Datasets and Their Characteristics (RQ1)}
\label{sec:results}

This section addresses our RQ1: an exploration of trends in open LA datasets. Throughout this section, the presented evidence pertains only to the datasets from the three conferences we investigated. Since these conferences represent the LA community, the findings capture the state of the art, but we cannot guarantee that they reliably generalize to the field as a whole. Therefore, the results should be interpreted in the context of LAK, EDM, and AIED.

\subsection{Executive Summary of the Observed Trends}
\label{subsec:results-summary}

Since \Cref{sec:results} is extensive, we provide a brief overview of the findings structured into easily readable bullet points. Details regarding each bullet point can be found in the corresponding referenced section. Each section also contextualizes the results and, when possible, compares them to prior work.

\Cref{subsec:results-A-distribution-of-papers} examines the selected papers that featured (meaning, provided or used) an open dataset:
\begin{itemize}
    \item Overall, \textit{\numselectedpapers\ of the \numcandidatepaperstext\ eligible papers} (19\%) featured an open dataset.
    \item Based on a relative count, \textit{EDM papers} featured open datasets the most, followed by AIED, and then LAK.
    \item \textit{Full papers} consistently featured open datasets more than short papers across all three venues.
    \item The featuring of open datasets \textit{slightly increased over the years}, but this trend was not statistically significant.
\end{itemize}

\Cref{subsec:results-B-usage} focuses on the dataset usage:
\begin{itemize}
    \item About \textit{a third of the \numselecteddatasets\ datasets were completely original}, while the remaining two-thirds were reused.
    \item \textit{LAK papers provided the most original open datasets}, followed by AIED, and then EDM.
    \item Most datasets were used once; however, \textit{20 were used three or more times}, with the most popular 12 times.
\end{itemize}

\Cref{subsec:results-C-availability} investigates the dataset availability:
\begin{itemize}
    \item Most selected datasets were \textit{freely available for download} -- the least restrictive option for the community.
    \item Of 35 datasets listed as on request, \textit{only half were provided} to us, demonstrating the unreliability of this option.
\end{itemize}

\Cref{subsec:results-D-educ-context} maps the educational contexts of the datasets:
\begin{itemize}
    \item The vast majority of datasets describe \textit{K-12 learners} (especially middle school students) and \textit{university students}.
    \item The vast majority of datasets are related to \textit{STEM education} (especially mathematics) and \textit{language learning}.
    \item Datasets from schools \textit{in the USA} dominate the research space.
\end{itemize}

\Cref{subsec:results-E-data-collection} examines the dataset collection:
\begin{itemize}
    \item More than a half of the datasets come from \textit{formal learning contexts} (for example, classrooms).
    \item Within those teaching contexts, the most prevalent teaching methods were \textit{quizzes, tests, and other questionnaires}.
    \item For collecting the data, \textit{system logging} and \textit{manual methods} are popular.
\end{itemize}

\Cref{subsec:results-F-size} looks at dataset sizes common for publication:
\begin{itemize}
    \item The sample size from which the data are collected varies widely, but the scale of \textit{hundreds of students} is typical.
    \item A typical dataset contains \textit{tens of thousands of data points}, although this information is highly context-dependent.
\end{itemize}

\Cref{subsec:results-G-technical} looks at technical properties:
\begin{itemize}
    \item The most popular hosting service for storing datasets is \textit{GitHub}.
    \item The vast majority of datasets have \textit{tabular format}, such as CSV.
\end{itemize}

Finally, \Cref{subsec:results-H-analytics} explores how datasets are analyzed in research papers:
\begin{itemize}
    \item \textit{Supervised machine learning} (potentially deep learning) is by far the most common analytical method.
    \item Popular metrics used for evaluating ML models in LA include \textit{accuracy, AUC, and F1-score}.
\end{itemize}

\subsection{Distribution of Papers That Feature an Open Dataset}
\label{subsec:results-A-distribution-of-papers}

\begin{table*}
\caption{Distribution of all the \numselectedpapers\ selected papers, that is, those that either used or provided an open dataset. The papers are again divided into full | short across the three target conferences throughout the years 2020--2024. In addition, the row and column \textbf{Total} include a percentage of the candidate papers compared to the corresponding cell in \Cref{tab:all-candidate-papers}.}
\label{tab:all-selected-papers}

\setlength\tabcolsep{4.8pt}

\begin{tabular}{l p{2mm} r|r p{2mm} r|r p{2mm} r|r p{2mm} r|r p{2mm} r|r p{2mm} r|r p{2mm} r}
\toprule
Full | Short & & \multicolumn{2}{c}{2020} & & \multicolumn{2}{c}{2021} & & \multicolumn{2}{c}{2022} & & \multicolumn{2}{c}{2023} & & \multicolumn{2}{c}{2024} & & \multicolumn{2}{c}{\textit{Subtotal}} & & \textbf{Total} \\
\midrule
LAK & &
4 & 0 & & 
6 & 4 & & 
1 & 1 & & 
7 & 1 & & 
10 & 4 & & 
\textit{28} & \textit{10} & & \textbf{38} (10.9\%) \\
EDM & &
9 & 7 & & 
6 & 13 & & 
6 & 7 & & 
9 & 7 & & 
7 & 11 & & 
\textit{37} & \textit{45} & & \textbf{82} (29.7\%) \\
AIED & &
6 & 10 & & 
7 & 7 & & 
12 & 5 & & 
13 & 4 & & 
15 & 5 & & 
\textit{53} & \textit{31} & & \textbf{84} (18.7\%) \\
\textit{Subtotal} & &
\textit{19} & \textit{17} & & 
\textit{19} & \textit{24} & & 
\textit{19} & \textit{13} & & 
\textit{29} & \textit{12} & & 
\textit{32} & \textit{20} & & 
\textit{118} & \textit{86} & & \textbf{\numselectedpapers} (19.0\%) \\
\textbf{Total} & &
\multicolumn{3}{l}{\textbf{36} (14.8\%)} & 
\multicolumn{3}{l}{\textbf{43} (17.8\%)} & 
\multicolumn{3}{l}{\textbf{32} (17.0\%)} & 
\multicolumn{3}{l}{\textbf{41} (22.7\%)} & 
\multicolumn{3}{l}{\textbf{52} (23.5\%)} & 
\multicolumn{3}{l}{\textbf{\numselectedpapers} (19.0\%)} \\
\bottomrule
\end{tabular}

\end{table*}

We discovered \numselectedpapers\ papers (19\% out of the \numcandidatepaperstext\ candidate papers) listed in \Cref{tab:all-selected-papers} that provided an open dataset or used a dataset that was already open. For simplicity, we group these two cases as \say{featuring an open dataset}. 

Compared to related work, \citet{sghir2023recent} reviewed 74 LA papers on predictive modeling, of which only three used an open dataset (OULAD in all cases). Our review had a broader scope and identified \numselectedpapers\ LA publications that employed an open dataset. \Cref{tab:all-selected-papers} implies several insights about these publications, which we detail below.

\subsubsection{Differences Between the Conferences}
The publication venues differed substantially in both absolute and relative counts. EDM featured open datasets more often than AIED, which in turn featured them more often than LAK. This finding could be explained by EDM's focus on computational data processing, whereas AIED balances data processing and pedagogy, and LAK focuses on pedagogy. In addition, since 2023, the calls for papers at EDM actively promote open science, encouraging the authors \say{to make their data, materials, and scripts openly available} whenever possible~\cite{EDM_CFP_open_science}.

\subsubsection{Differences Between the Paper Types}
Full papers were more likely to feature an open dataset (118 of 571 candidate papers, 20.7\%) compared to short papers (86 of 505 candidate papers, 17.0\%). When looking at relative counts, this trend was also preserved at each of the three individual conferences. At LAK, this was 12.2\% vs. 8.3\%; at EDM, 31.9\% vs. 28.1\%; and at AIED, 23.6\% vs. 13.8\%. Even though EDM had a larger absolute count of short papers with datasets than full papers, the relative percentages still favor full papers.

\subsubsection{Differences Between the Individual Years}
Although five years might be too short a time frame to observe a clear trend, there was an overall indication that the number of open datasets in the published LAK, EDM, and AIED papers increases over the subsequent years (except for a slight decrease from 2021 to 2022).

To validate our assumption, we performed a trend analysis using the Mann-Kendall test~\cite{mann1945nonparametric}, which was also used in other review papers~\cite{chen2020detecting, Chen2022_two_decades_AIED}. The test's null hypothesis is that \textit{throughout the five years, there is no significant monotonically increasing/decreasing trend in the number of open datasets at the three conferences}. However, for each conference, the resulting p-value was larger than the 0.05 threshold, failing to reject the null hypothesis, with a small positive Sen's slope effect size ($datasets/year$) for AIED ($S = 1.25$, 95\% CI [-2.0, 3.0]), EDM ($S = 0.25$, 95\% CI [-6.0, 3.0]), and LAK ($S = 1.92$, 95\% CI [-8.0, 6.0]). Therefore, the small upward trend we observed intuitively may not be significant. 

We also used sensitivity analysis \cite{saltelli2002making} to examine the effects of design choices, such as the treatment of on-request datasets and the 30-day response window, on the trend analysis results. Papers excluded based on these criteria were included in the dataset frequency as upper bounds for the parameter. The results from 1024 generated samples indicated that while adding excluded datasets showed first-order sensitivities in select years, higher-order interactions were minimal, suggesting these design choices did not affect significance testing in the trend analysis. This implies that further awareness of the benefits of open datasets in LA is needed, underscoring the importance of our survey.

\subsection{Usage of the Datasets}
\label{subsec:results-B-usage}

The \numselectedpapers\ selected papers employed a total of \numselecteddatasets\ unique datasets -- see the full list in \Cref{sec:appendix-datasets}, \Cref{tab:datasets}. 

\subsubsection{Original Datasets}
Out of the \numselectedpapers\ selected papers, 56 papers provided 56 fully original datasets (that is, appearing for the first time in that paper). These 56 original datasets represent 32.6\% of all of the \numselecteddatasets\ unique datasets, while the remaining 116 datasets (67.4\%) were reused from a previously published location. 

Of the 56 original datasets, 16 were published alongside LAK papers (42.1\% of the 38 selected), 17 were published with EDM papers (20.7\% of the 82 selected), and 23 with AIED papers (27.4\% of the 84 selected). Surprisingly, this is the reverse of the trend observed above: although LAK papers were the least likely to feature an open dataset overall, they were the most likely to provide an original, novel dataset. At the same time, EDM focused most on reusing existing open datasets rather than creating new ones. AIED again remained in the middle of the other two conferences.

\subsubsection{Multiplicity of Dataset Use}
As a reminder, there was an \textit{m:n} relationship between the \numselectedpapers\ selected papers and \numselecteddatasets\ datasets -- that is, one paper could have used more datasets, and one dataset could have been used in multiple papers.

In total, there were 286 occurrences of open dataset use: 47 uses across the 38 LAK papers, 121 uses across the 82 EDM papers, and 118 uses within the 84 AIED papers. This means that on average, each selected LAK paper used 1.24 datasets, each selected EDM paper used 1.48 datasets, and each selected AIED paper used 1.40 datasets. Again, this makes LAK papers use the least open datasets per paper, EDM the most, and AIED in the middle.

From the perspective of full vs. short papers, the 286 occurrences of open dataset use were distributed across 164 in full papers (57.3\%) and 122 in short papers (42.7\%), a relatively balanced distribution. 

\subsubsection{Repeated Use of the Same Dataset}

Of the \numselecteddatasets\ unique datasets, 132 were used only in a single paper (76.7\%), 20 were used in two papers (11.6\%), and the remaining 20 were used in three or more papers (11.6\%). For each dataset, its usage count and the papers that used it are listed in \Cref{tab:datasets}, with the 20 most \say{popular} datasets highlighted.

Among these 20 datasets, 4 belong to the ASSISTments system~\cite{ASSIST}. 
Next, 2 other datasets have won the EDM dataset prize~\cite{EDM_dataset_prize} -- NeurIPS 2020 (ID 103 in \Cref{tab:datasets}) and OULAD (ID 109), both used 6 times each. However, another EDM dataset prize winner (ID 12) was used only once, indicating that the winning datasets are not necessarily reused.
Finally, the remaining reused datasets have often been used in LA competitions, such as the ASAP prize (IDs 5 and 6).

The observation that relatively few datasets are used very often matches prior reviews. In a recent survey of EDM open science practices, \citet{Baker2024open} observed a similar phenomenon in knowledge tracing, noting that \say{a small set of datasets are utilized in the super-majority of recent papers on knowledge tracing}.

\subsection{Availability of the Datasets}
\label{subsec:results-C-availability}

Our second selection criterion (see \Cref{subsec:methods-criteria}) defined the rules for obtaining the dataset electronically. Mirroring this criterion: (a) 130 selected datasets (75.6\%) were available by direct download, (b) 24 selected datasets (14.0\%) were available for download after logging in, and (c) the remaining 18 selected datasets (10.5\%) were available on request.

In addition, 17 datasets were listed in the corresponding paper as available on request, but were not provided to us by the respective owners. The reasons, sorted by the most common, were:
\begin{itemize}
    \item not responding to our request (12\texttimes), 
    \item declining our request (3\texttimes), 
    \item taking more than one month to respond (1\texttimes), and
    \item inability to contact the authors, as all their listed email addresses were no longer functional (1\texttimes).
\end{itemize}
Due to eliminating these 17 datasets, we excluded an additional 16 papers that would otherwise have been selected.

Overall, of the 35 datasets claimed to be available on request, we had to eliminate half. We speculated that, in some cases, the dataset owner had moved on to another position, making them difficult to reach. In other cases, sophisticated procedures were required to obtain the dataset (for example, signing a license agreement and sending it to the authors); yet, even though these procedures took a long time to complete, no response or follow-up was provided.

This issue is also prevalent in related research: as early as 2005, \citet{wicherts2006poor} reported on \say{The poor availability of psychological research data for reanalysis}. Of 249 requested datasets from recent studies, they received only 64 (25.7\%). Although our success rate was higher (18 out of 35, or 51.4\%), this still means that if LA researchers request data (from a paper stating that such a request is possible), the success chance is similar to a coin toss.

Mirroring \citet{Tedersoo2021}, who claimed that \say{statements of data availability upon (reasonable) request are inefficient and should not be allowed by journals} -- a guideline now adopted by large publishers such as \citet{open-data-licenses-nature} -- we doubt whether sharing datasets on request is suitable for the LA community. The other two options (direct download or download after logging in, if security steps are needed) are more suitable in most cases.

\subsection{Educational Contexts of the Datasets}
\label{subsec:results-D-educ-context}

In LA research, the educational context of data collection is crucial for subsequent interpretation of the data analysis. Therefore, we examined various aspects pertaining to the educational context.

\subsubsection{Level of Students Represented in the Dataset}
Although some LA datasets may pertain to teachers or educational administrators, most of them describe students. Across our \numselecteddatasets\ selected datasets, the most common categories were\footnote{Since multiple categories can be associated with one dataset, the sum of occurrences is larger than the count of all datasets.}:
\begin{itemize}
    \item K-12 students (94\texttimes), divided into: high school (28\texttimes), middle school (40\texttimes), elementary school (11\texttimes), and otherwise unspecified K-12 context (15\texttimes).\footnote{Educational systems in different countries may use different divisions between high, middle, and elementary school. Therefore, this division is not always 100\% comparable, although the differences are usually minor.}
    \item University students (55\texttimes).
    \item Undefined, unspecified, or otherwise unclear educational level (36\texttimes, which is rather often).
    \item Not applicable (12\texttimes, such as in the case of simulated datasets).
\end{itemize}
The other, more rare categories were: professional learners (8\texttimes), various learners across all levels grouped together (4\texttimes), and PK-12 (1\texttimes). 

As expected, K-12 and university contexts dominate, given the large number of students at these levels and the contexts' broad definition spanning several years. At the same time, this presents a gap, as the LA community lacks enough open datasets to examine other groups of learners. 

\subsubsection{Educational Topics}
The datasets are related to various teaching topics. To avoid bias towards a specific country's curriculum, we categorized the topics according to UNESCO's International Standard Classification of Education (ISCED)~\cite{ISCED_categories}. For simplicity, we list only the topics that occurred 5 or more times.

\begin{itemize}
    \item By a large margin, the most common field was \say{05 Natural sciences, mathematics and statistics}, with 83 occurrences total -- divided into 52\texttimes\ mathematics and statistics, 15\texttimes\ on physical sciences (physics, chemistry, earth sciences), 8\texttimes\ on biology and biochemistry, and 8\texttimes\ further unspecified.
    \item The second most common field was \say{02 Arts and humanities}, with 36 occurrences -- 32 on languages, 3 on humanities (except languages), and 1 on arts.
    \item The third most common field was \say{06 Information and Communication Technologies} with 29 occurrences, especially in the computing subfields of programming and data science.
    \item There were 21 datasets with a mix of topics spanning more than three categories.
    \item Next, there was engineering (6\texttimes) and health (6\texttimes).
    \item Lastly, 5\texttimes\ each: basic literacy and numeracy, social and behavioral sciences, and undefined.
\end{itemize}

Since ISCED defines 10 broad educational fields, and the datasets focused on STEM-related areas, there is an untapped opportunity for researchers to explore and publish datasets from other educational contexts. Areas that were insufficiently covered include, but are not limited to, education in the social sciences, business, and agriculture.

Compared to related work by \citet{Svabensky2022applications}, who reviewed 35 LA papers in cybersecurity education, our survey did not identify any paper whose educational context would focus on cybersecurity. Among the computing topics (\say{06 Information and Communication Technologies}), programming and data science were the prevalent themes. 

\begin{figure}[t]
\centering
\includegraphics[width=0.64\linewidth]{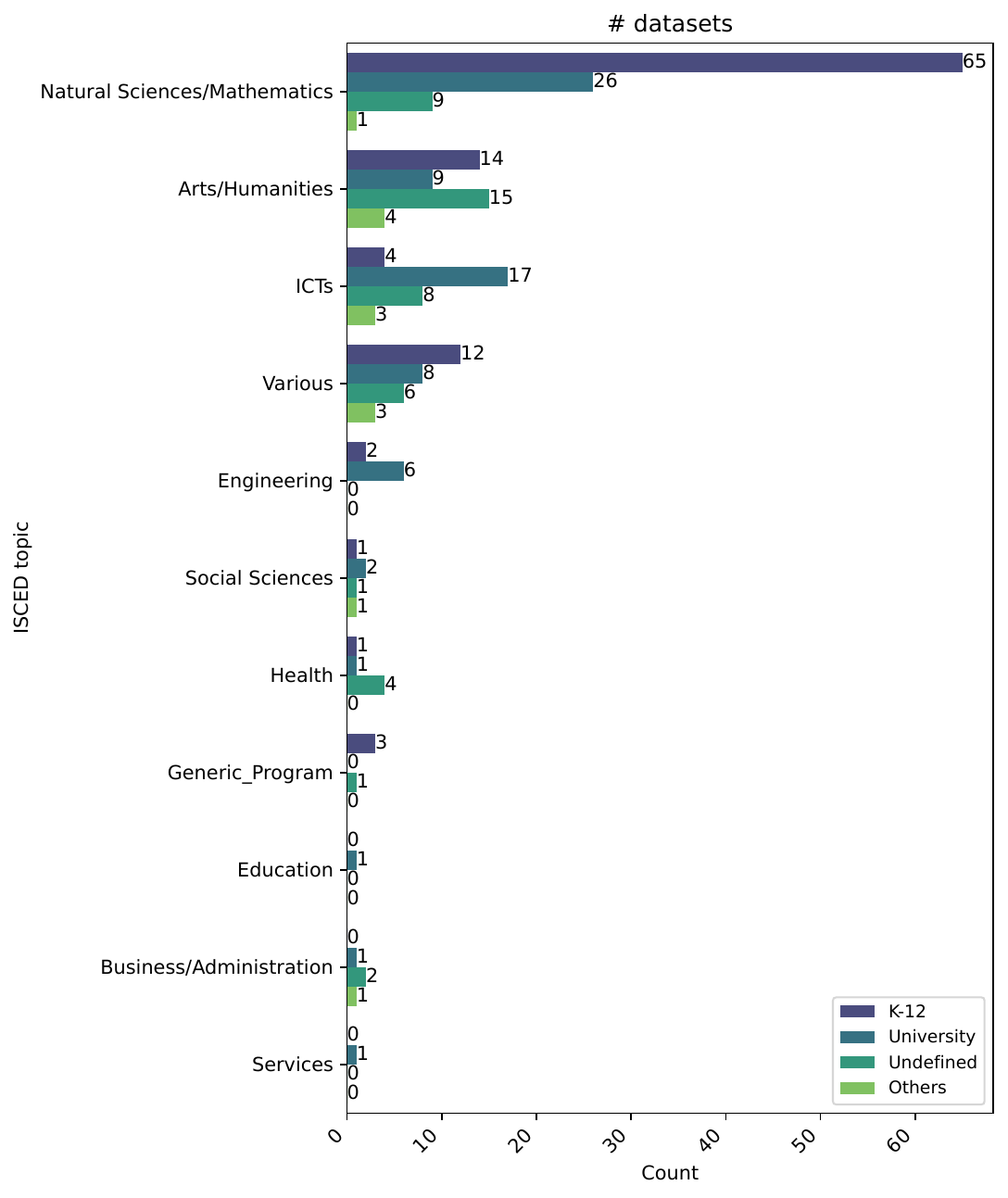}
\caption{Distributions of dataset frequency across educational topics and levels of students.}
\Description{Bar chart showing the prevalence of natural sciences topics and K-12 contexts.}
\label{fig:topics-students}
\end{figure}

Cross-referencing the educational topics with the above-mentioned level of students, \Cref{fig:topics-students} visualizes the distribution of datasets. The strong presence of datasets from natural sciences education of K-12 and university students is prevalent. At the same time, the figure highlights the underdeveloped categories.

\subsubsection{Geographical Distribution}
\begin{figure}[t]
\centering
\includegraphics[width=\linewidth]{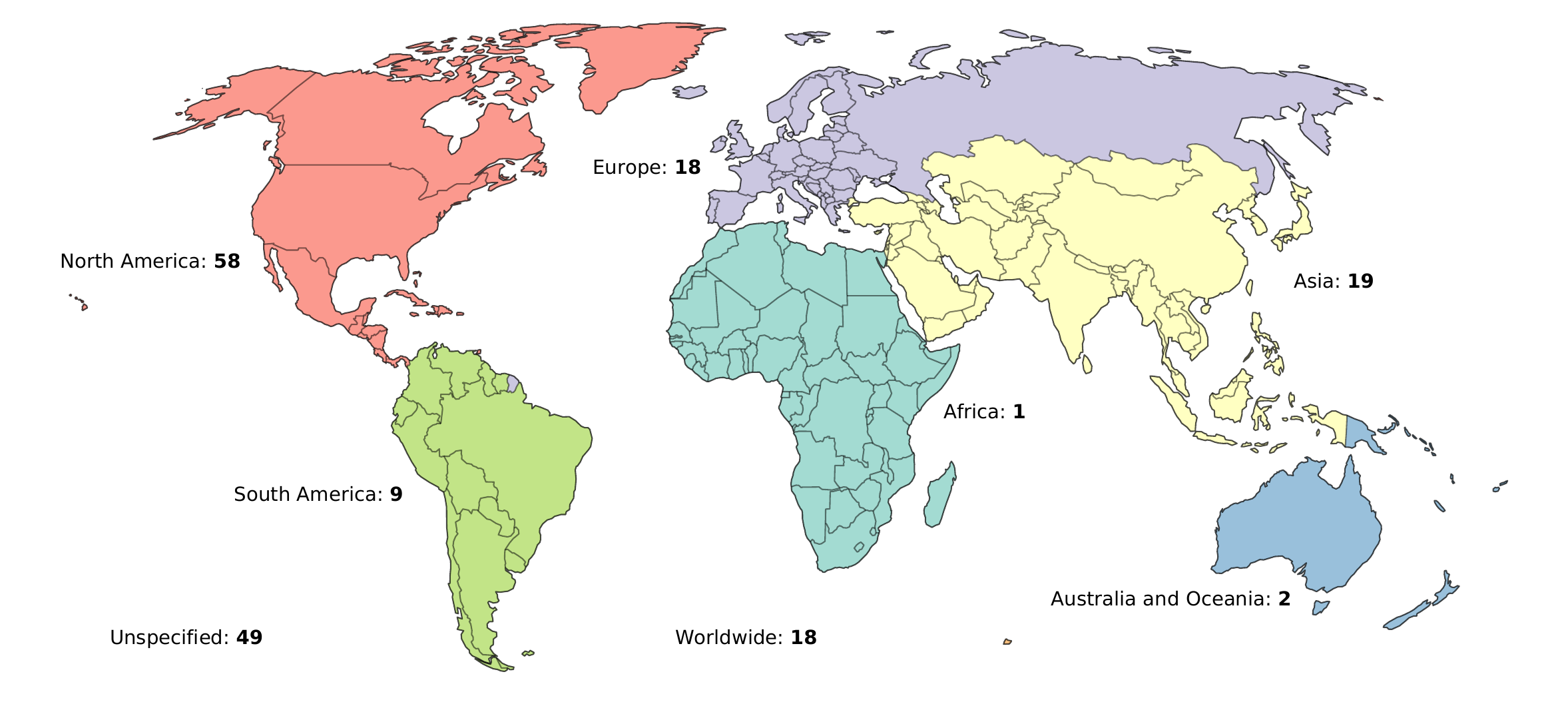}
\caption{Geographical contexts of the dataset collection distributed per continent. This map considers 163 datasets (\numselecteddatasets\ minus the 9 that are not applicable). However, some datasets pertained to multiple countries/continents, so the counts in the map sum to 174. The map was automatically generated in Python using the \texttt{geopandas} module~\cite{GeoPandas}.}
\Description{World map with highlighted continents, showing the prevalence of dataset origin in North America.}
\label{fig:continents}
\end{figure}

The location where the dataset was collected is another interesting factor, shown in \Cref{fig:continents}. Of the \numselecteddatasets\ datasets, 49 (that is, 28.5\%) did not indicate the country or even the continent of origin. For 9 datasets, this information was not applicable (7 synthetic/simulated and 2 consisting solely of math questions). Among the remaining 114, 18 were \say{worldwide} with users globally, such as from apps like Duolingo. As for the remaining 96 datasets, 92 pertained to exactly one continent, while 4 combined data from two or more continents. North America had a larger presence than all other continents combined. Next, these dataset origin countries were present more than once:
\begin{itemize}
    \item \textit{Most prevalent:} USA (56\texttimes),
    \item \textit{More than 3\texttimes :} China (7\texttimes), Brazil (5\texttimes), and India (4\texttimes),
    \item \textit{3\texttimes\ each:} France, Germany, Japan, and UK,
    \item \textit{2\texttimes\ each:} Colombia, Ireland, South Korea, and Switzerland. 
\end{itemize}

The dominance of US-based datasets reflects LA's research focus in the North American context. This presents opportunities for underrepresented regions to contribute to the dataset space. \citet{Zheng2022} observed that a few Global South institutions were represented at LAK, EDM, and AIED; we corroborate this trend regarding datasets.

\subsection{Data Collection Properties of the Datasets}
\label{subsec:results-E-data-collection}

This section expands on \Cref{subsec:results-D-educ-context} regarding the teaching context, but focuses specifically on data collection. 

\subsubsection{Teaching Environments}
All \numselecteddatasets\ datasets were distributed as follows:
\begin{itemize}
    \item 94 were collected strictly within formal learning contexts, for example, in classrooms and university courses.
    \item 11 were collected during informal learning, such as extracurricular learning, and 1 combined both formal and informal contexts.
    \item 26 were collected in an artificial setting, such as a controlled environment during a lab study.
    \item 9 datasets were collected from non-educational contexts, such as annotated social network comments.
    \item Finally, for 13 datasets, this information was not applicable, and 18 did not specify the environment.
\end{itemize}

\subsubsection{Teaching Methods}
Regarding the teaching/learning methods applied during the data collection (some of which could have been combined for one dataset), the most common were:
\begin{itemize}
    \item quizzes and questionnaires (66\texttimes),
    \item hands-on exercises and assignments other than writing (36\texttimes), 
    \item tutoring systems (25\texttimes), 
    \item writing tasks (17\texttimes),
    \item MOOCs (17\texttimes),
    \item e-learning platforms (10\texttimes), 
    \item lectures (9\texttimes), and
    \item for 17 datasets, this information was not applicable, and 5 did not specify the method.
\end{itemize}

\subsubsection{Method of Data Collection}
Regarding the data collection instruments (which could also be combined, like in the previous case), the most common were:
\begin{itemize}
    \item logs of system actions (76\texttimes),
    \item manual data collection (68\texttimes), 
    \item recording questionnaire answers (35\texttimes), 
    \item web scraping (23\texttimes), 
    \item transcripts of speech (10\texttimes), 
    \item simulations (9\texttimes), 
    \item biometrics (4\texttimes), and
    \item 4 datasets did not specify the method.
\end{itemize}

\subsection{Sizes of the Datasets}
\label{subsec:results-F-size}

We examined two aspects of the dataset size: the number of students represented in the dataset and the number of data points. While the first metric is comparable across datasets, the second metric depends on the dataset type. Still, both are useful for providing an overview of the typical dataset sizes reported in LA research.

During data collection, we sometimes encountered discrepancies between the source paper describing the dataset and the dataset itself. For example, a paper reported that the dataset has 1,000 students, but the dataset file contained 990 unique student IDs. In cases like these, we recorded the actual numbers in the dataset within our spreadsheet.

\subsubsection{Number of Students}

\begin{figure}[t]
\centering
\includegraphics[width=0.8\linewidth]{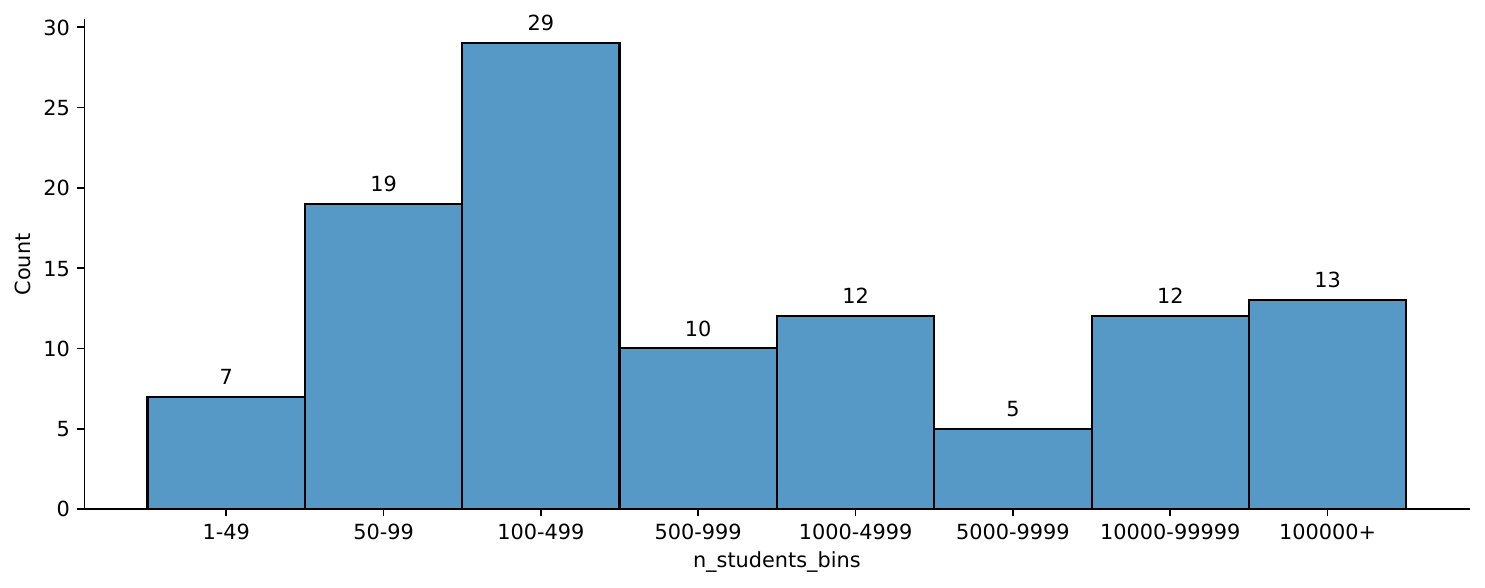}
\caption{Distribution of number of students into bins.}
\Description{Bar chart showing the various categories of the number of students.}
\label{fig:num_students}
\end{figure}

This information applied to 130 out of all \numselecteddatasets\ datasets (75.6\%). The remaining ones did not collect data about students, but, for example, about courses or teaching materials. Of the 130 applicable datasets, 23 (17.7\%) do not disclose this information. The remaining 107 datasets are distributed in \Cref{fig:num_students}. The most common category is 100--499 students, which seems reasonable for LA research, as datasets about hundreds of students can show statistically significant effects or result in better-performing ML models. Surprisingly, even huge datasets (over 10,000 students) are sufficiently represented. The median sample size across these 107 datasets is 478 students.

To provide context, we compare this number with a prior review of 35 LA papers on cybersecurity education~\cite{Svabensky2022applications}. The median sample size for data analytics was 43 students. Within the general LA datasets identified in the present survey (none of which were related to cybersecurity education), the median number of students was much larger.

\subsubsection{Number of Data Points}

\begin{figure}[t]
\centering
\includegraphics[width=\linewidth]{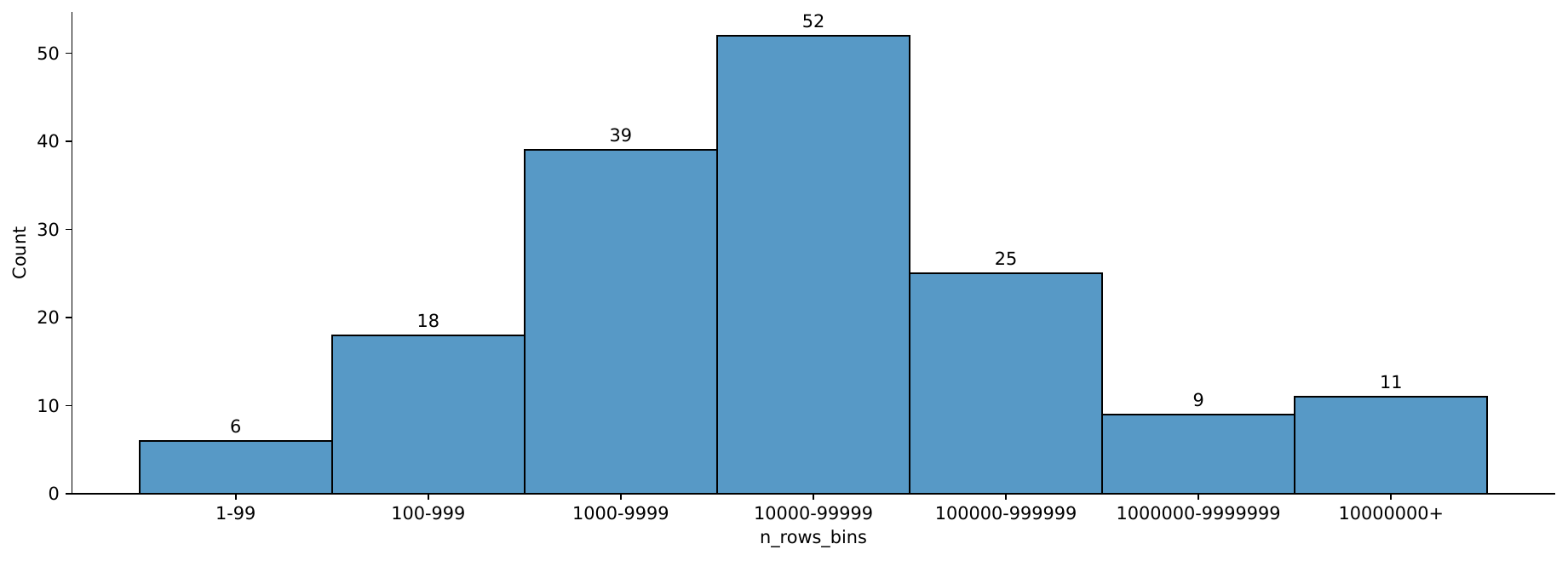}
\caption{Distribution of number of data points into bins.}
\Description{Bar chart showing the various categories of numbers of data points.}
\label{fig:num_data_points}
\end{figure}

This information was applicable to all \numselecteddatasets\ datasets, although 12 of them (7.0\%) did not provide this information, and it was not feasible for us to determine it without unreasonable effort. The remaining 160 datasets are distributed in \Cref{fig:num_data_points}. The median size across these 160 datasets is 14,457 data points.

Compared to prior work, \citet{Stuart2018} observed that in non-LA domains, small datasets ($<$ 20 MB) are shared substantially less than large datasets ($>$ 50 GB). Although we did not examine file size, the number of data points strongly correlates with it. However, we did not observe any such trend. \Cref{fig:num_data_points} has a normal-like distribution and shows that very small and very large datasets are rarely shared, while medium-sized datasets are prevalent.

\subsection{Technical Properties of the Datasets}
\label{subsec:results-G-technical}

\subsubsection{Hosting Services to Store Datasets}
We observed the following dataset repositories used by authors.
\begin{itemize}
    \item By far the most popular storage location was GitHub (64\texttimes, 37.2\%). In contrast, GitLab was used only once. 
    \item Second came PSLC DataShop (27\texttimes, 15.7\%), and OSF was third (12\texttimes, 7.0\%).
    \item Following were Kaggle (9\texttimes), Google Drive (6\texttimes), Google Sites (6\texttimes), and Zenodo (5\texttimes).
    \item The remainder were different websites, mostly institutional or personal, each occurring only once or twice. 
\end{itemize}

\subsubsection{File Formats to Store Datasets}
The formats of the dataset files also varied.
\begin{itemize}
    \item By far, the most popular format was CSV (108\texttimes), which can be grouped with TSV (9\texttimes).
    \item Next was TXT (36\texttimes),
    \item followed by JSON (21\texttimes) with JSONL (5\texttimes),
    \item and XLSX (14\texttimes) with XLS (2\texttimes).
    \item Finally, there was XML (4\texttimes), DAT (3\texttimes), and DB (3\texttimes).
    \item The remaining formats were used very rarely, only once or twice.
\end{itemize}

\subsection{Analytics of the Datasets Within the Selected Papers}
\label{subsec:results-H-analytics}

While the previous sections dealt with the \numselecteddatasets\ datasets themselves, we now focus on how the \numselectedpapers\ selected \textit{papers} analyzed the datasets. All these papers reported some form of data analytics, except one, which focused solely on describing the dataset.

\subsubsection{Analytical Methods and Metrics}

We base this section on \citet{Aldowah2019}, who reviewed data analysis techniques used in LA publications, which we subdivided into categories:
\begin{itemize}
    \item Supervised machine learning -- classification, regression,
    \item Unsupervised machine learning -- clustering, outlier detection,
    \item Data mining -- association rule mining, sequential pattern mining, text mining, visual data mining,
    \item Statistical techniques -- descriptive and inferential statistics, correlation/causal mining, density estimation.
\end{itemize}
We used these categories as a baseline and extended them to define 10 analytics categories, listed in \Cref{tab:analytics}, sorted by frequency. The goal is to help readers identify which metrics they might consider based on the type of analytics they plan to conduct. We note that one paper often used more than one type of analytics, so the summary counts exceed the total of \numselectedpapers\ papers. Also, the categories may overlap, for example, knowledge tracing is a part of supervised ML~\cite{Abdelrahman_KT_survey}, but it occurred so often that we listed it as a separate category.

\begin{table}[t]
\caption{The analytical categories and their most common metrics used with the datasets across the \numselectedpapers\ selected papers.}
\label{tab:analytics}

\rowcolors{2}{gray!15}{white}
\begin{tabular}{rll}
\toprule
\textbf{Usage} & \textbf{Analytic} & \textbf{Common metrics (sorted by usage)} \\ 
\midrule

137\texttimes & Supervised ML & accuracy (50\texttimes), AUC (45\texttimes), F1 (38\texttimes), RMSE (19\texttimes), recall (17\texttimes), precision (13\texttimes) \\ 

102\texttimes & Deep learning & accuracy (34\texttimes), AUC (30\texttimes), F1 (30\texttimes), recall (12\texttimes), RMSE (10\texttimes), MAE (10\texttimes) \\ 

 75\texttimes & Statistics & inferential tests (30\texttimes), correlations (19\texttimes) \\ 

 38\texttimes & Knowledge tracing & AUC (25\texttimes), accuracy (13\texttimes) \\ 

 17\texttimes & Large language models & F1 (7\texttimes), accuracy (6\texttimes) \\ 

  8\texttimes & Unsupervised ML & custom metrics, such as cluster properties \\ 

  5\texttimes & Data mining & custom metrics depending on the technique \\ 

  3\texttimes & Reinforcement learning & accuracy (1\texttimes) \\ 

  2\texttimes & Visualization only & N/A \\ 

  2\texttimes & Other context-specific & custom metrics depending on the technique \\ 

\bottomrule
\end{tabular}

\end{table}

As for the evaluation metrics, we followed up on \citet{Bai2021metrics-review}. They list commonly used metrics in LA: accuracy, precision, recall, F1-score, AUC, MAE, and RMSE, to which we added more metrics as necessary during our survey.

\subsubsection{Sharing of Analytical Tools}

Of the 202 papers that analyzed a dataset, 96 (47.5\%) provided a public URL linking to the analytical tools or software used in the paper. This is commendable, as it facilitates research reproducibility and replicability. However, we did not test the deployment of these tools, as this is outside the scope of this survey. We only reviewed the storage URLs: GitHub was the most popular service for publishing research code (86\texttimes), followed by OSF (8\texttimes), and finally GitLab (1\texttimes) and Google Drive (1\texttimes).

\section{Implications and Conclusions}

Compared to prior work~\cite{romero2020, Mihaescu2021review, bigdata2021edudata, lin2024surveydeeplearning, Xiong2024survey}, our survey has identified the largest number of open datasets (\numselecteddatasets) for LA research, along with \numselectedpapers\ associated publications. We also revealed the most detailed insights about these datasets and the publications that used them, summarized in \Cref{subsec:results-summary} (which addresses our RQ1).

In this part, our paper provides practical implications to LA researchers and practitioners:
\begin{itemize}
    \item \Cref{subsec:implications-future-work} reflects on the open challenges in the field, resulting from our analysis of trends.
    \item \Cref{subsec:implications-recommendations} provides guidelines for LA data sharing, summarized as a \say{PRACTICE} checklist in \Cref{sec:appendix-checklist}.
    \item \Cref{subsec:implications-materials} describes the openly available data and materials we provide alongside our research, implementing our checklist and demonstrating the outcome in practice.
\end{itemize}

\subsection{Open Research Challenges and Opportunities for Future Work (RQ2)}
\label{subsec:implications-future-work}

The trends identified in \Cref{sec:results} reflect the state of the art in LA research conducted with public datasets. Although we excluded papers with private datasets, by identifying \numselectedpapers\ papers, we captured a substantial snapshot of the current LA research space. We now discuss the issues the community cannot explore with open data alone, because such data are not available. We hope to inspire other researchers to address some of the open challenges identified:
\begin{enumerate}
    \item While K-12 learners and university students are often represented, much less is known about LA applications concerning other stakeholders, such as professional learners in the workplace or teachers.
    \item Most open LA datasets focus on students who learn STEM topics or languages. Many other educational fields were not explored much in prior work.
    \item The vast majority of open datasets were collected in the USA -- more than in all other countries combined. This represents an opportunity for other regions to contribute to the LA dataset space.
    \item Methods other than supervised machine learning can be explored further. These include unsupervised learning, reinforcement learning, language models, or other less traditional approaches to deliver educational analytics.
    \item Throughout 2020--2024, we did not observe a significant increasing trend in the number of datasets made available. This indicates a need for a stronger push to make datasets open, which our paper supports by providing actionable guidelines (\Cref{subsec:implications-recommendations}) and a \say{PRACTICE} checklist (\Cref{sec:appendix-checklist}).
\end{enumerate}

We also identified four types of future survey papers relevant to the LA community.
\begin{enumerate}
    \item The most straightforward way to extend this survey is to explore more data sources. These include field-specific journals (for example, the Journal of LA, the Journal of EDM, and the International Journal of AIED), as well as journals focused on research data (for example, Springer Scientific Data and Elsevier Data in Brief). In the latter case, our basic keyword search for \say{learning analytics} revealed several datasets published in these data-focused journals. Although there are not many examples yet, those that exist~\cite{Svabensky2021dataset, Chytas2023, chen2024accoding} demonstrate that LA is broadly relevant beyond the scope of the three specific communities. 
    \item Future work can extend our methods by evaluating openness quality. It can adopt a brief rubric to score datasets on access mode, licensing, documentation, metadata, persistent identifiers, versioning, privacy safeguards, repository stability, and code and citation guidance.
    \item Another interesting survey would be to examine papers that claim that LA needs more research on a certain topic $X$, and investigate whether relevant data exist to do that $X$. This would help the research community focus its attention on problems that are yet unresolved and can be tackled with the available datasets.
    \item Understanding the situations when a dataset cannot be provided is also valuable. Relevant future work would be a global survey of potential barriers to data sharing by country of origin. This would provide an overview of legal and practical limitations and differences across the world.
\end{enumerate}

\subsection{Actionable Recommendations and Guidelines to Support Researchers (RQ3)}
\label{subsec:implications-recommendations}

In 2019, Springer Nature identified the following five essential factors to facilitate data sharing~\cite{Lucraft2019}:
\begin{enumerate}
    \item \textit{Proper credit} -- researchers want to receive sufficient credit, and they should be formally recognized through co-authorship, citations, better research assessment, and points towards career advancement (tenure, grants).
    \item \textit{Clear policy} -- institutions, publishers, and communities should set specific requirements for data sharing.
    \item \textit{Explicit funding} -- to ensure policy compliance, funding institutions should provide dedicated financial support.
    \item \textit{Practical help} -- researchers should receive guidance with organizing data, selecting suitable repositories, and other steps of the process to make it as simple and efficient as possible.
    \item \textit{Training and education} -- compile data sharing best practices, communicate benefits, and address concerns.
\end{enumerate}

In the field of psychology, which is related to LA from the educational perspective, \citet{Meyer2018_tips_data_sharing} provided tips for \say{do's} and \say{don'ts} of data sharing:
\begin{itemize}
    \item \textit{Don't promise to}: (1) destroy data, (2) not share data, or (3) limit data analysis to certain topics.
    \item \textit{Do}: (4) get consent to retain and share data, (5) perform anonymization, and (6) select a suitable data repository. These tips align with what \citet{Baker2024open} recommends for the LA context.
\end{itemize}

Based on our survey findings, we complement these guidelines with eight practical recommendations for LA researchers. We also aim to clarify and address some of the concerns regarding dataset sharing presented in \Cref{sec:intro}. Our eight recommendations are detailed in this section and then summarized in \Cref{sec:appendix-checklist} under the acronym PRACTICE. Our guidelines are orthogonal to the well-established \say{DELICATE} checklist for trusted LA~\cite{Drachsler2016DELICATE}.

\subsubsection{Prepare for Open Data Sharing Before Collecting the Data}
As \citet{Baker2024open} pointed out, the possibility of data sharing must be considered from the beginning. It needs to be reflected in the application sent to the institutional review board (IRB, also called the ethical committee at some institutions), and the research participants must be aware of it when signing research consent agreements~\cite{Kathawalla2021easing}.

Thus, before data collection begins, it is essential to prepare for the later stages of data anonymization (that is, de-identification). This measure should be communicated to the ethical committee and the research participants.
\begin{itemize}
    \item[$\checkmark$] In the IRB application, explain that the research data will be \textit{anonymized} (de-identified).
    \item[$\checkmark$] In the research consent form, assure the participants that their data will be \textit{anonymized} (de-identified).
\end{itemize}

In addition, prepare for the option to share the dataset in the future.
\begin{itemize}
    \item[$\checkmark$] In the IRB application, explain that after the precautions, the data may be publicly \textit{shared} for research purposes.
    \item[$\checkmark$] In the consent form, ask the participants for permission to \textit{share} the anonymized data for research purposes.
\end{itemize} 

Sometimes, the specific research may not require IRB approval under local regulations (as is the case with this survey, which does not involve human participants and uses only publicly available data); this should be stated in the article.

Lastly, there are cases when the dataset or parts of it cannot be released. One example is when the research requires collecting sensitive data or data that cannot be completely anonymized. Another example is preparing for the possibility that a student may request that their data be deleted, as thoroughly discussed by \citet{hutt2023right}. Further examples were listed in \Cref{subsubsec:related-work-cautionary}. In cases like these, the ethical code of conduct for research and legal regulations take priority. Still, if there is a subset of the dataset (for example, only certain variables) that can be shared, researchers can prepare to share only the data they have the right to share. If data cannot be shared for external reasons beyond the authors' control, it would be helpful to explain the barriers in the associated article. 

In the rest of this section, our guidelines will assume that data can be shared.

\subsubsection{Refine and Preprocess the Data for Open Use}
After collecting the dataset, thoroughly anonymize it as promised. To avoid the risk of re-identification~\cite{Kathawalla2021easing}, this involves removing personally identifiable information (for example, student names)~\cite{bosch2020hello} and potentially sensitive attributes such as demographics~\cite{Baker2023demographics, fEDM23investigating}. After all, demographic data are no longer recommended for use in typical LA applications, such as predictive modeling~\cite{Baker2023demographics, fEDM23investigating}.
\begin{itemize}
    \item[$\checkmark$] Perform anonymization by removing personal information associated with human participants.
\end{itemize}

Alternative anonymization methods, such as synthetic data generated based on the analysis of original datasets' properties, are also increasingly being explored to enable dataset sharing~\cite{flanagan2022fine}. This can be a suitable alternative to releasing the original dataset when only derivative data may be released.

Although tools such as LLMs can help automate the anonymization task~\cite{Singhal2024deidentifying}, they are still error-prone, so human oversight is essential. We recommend having at least one person responsible for the anonymization and at least one other person not involved in the process to perform independent validation.
\begin{itemize}
    \item[$\checkmark$] Have an independent human validator to double-check the dataset contents for any anonymization issues.
    \item[$\checkmark$] In the dataset description, ask the future users to report to you should they find any anonymization issues.
\end{itemize}

\subsubsection{Add Clear Documentation and Metadata}
A missing data dictionary or other documentation can severely limit others' ability to use the dataset. For example, in tabular datasets (by far the most common type), the data dictionary is a file that describes the meaning of individual columns and their possible values. The ASSISTments2017 dataset (ID 22 in \Cref{tab:datasets}) includes an example of good practice. However, within the \numselecteddatasets\ datasets we discovered, 62 of them (36\%) lacked this type of documentation. In line with \citet{Kathawalla2021easing}, we strongly recommend describing the dataset structure for future users.
\begin{itemize}
    \item[$\checkmark$] Include the data dictionary alongside the dataset to enable other researchers to use your dataset.
\end{itemize}

In addition, the associated properties, such as the educational aspects (teaching topic, data collection location) and size (number of students and data points), were often missing in the \numselecteddatasets\ datasets. Such information helps users better understand the dataset's context and the potential limitations of its analytics.
\begin{itemize}
    \item[$\checkmark$] Describe the properties and contextual information of the dataset and its variables.
\end{itemize}

\subsubsection{Choose a Trusted and Stable Repository for Longevity}
As explained by \citet{Meyer2018_tips_data_sharing}, using a suitable data repository is essential to ensure long-term archiving and data accessibility. However, \Cref{sec:intro} discussed that choosing a repository was the third largest \say{pain point} for researchers, with 33\% being unsure about which repository to use~\cite{Stuart2018}. We address this concern below.

Across the \numselecteddatasets\ surveyed datasets, GitHub was by far the most popular. Although this is a solid choice, as many researchers and practitioners are familiar with it, dedicated research repositories may offer more stability. For example, researchers in the United States may prefer OSF, and Zenodo is popular within the European Union. Both options assign a persistent identifier (DOI) to the dataset, which is a significant advantage for longevity. In addition, more and more publishers have now been offering the option to host supplementary materials alongside the published article.

We recommend Zenodo due to its focus on long-term preservation on CERN's infrastructure, easy DOI-based versioning with automated GitHub integration, and allowing a wide variety of file formats and sizes. Zenodo also provides standardized metadata fields and displays the tree-like structure of dataset subfolders, which helps with navigation and organization -- the largest \say{pain point} reported by 46\% of researchers~\cite{Stuart2018}.
\begin{itemize}
    \item[$\checkmark$] Use a well-known, long-term storage for the dataset, such as Zenodo.
    \item[$\checkmark$] Ensure the data files follow a standard format to ensure long-term readability. 
    \item[$\checkmark$] Structure the dataset files so they are understandable to outside users. For example, use a division into suitably named subfolders that are logical for your specific context, and describe them in a Readme file.
\end{itemize}

Other suitable alternatives to Zenodo include HuggingFace, which offers robust data-sharing integration, OSF, or institutional GitHub. Regardless of which of these services are used, the existing infrastructure and upload templates save time and costs -- the fourth and fifth \say{pain points} identified by 26\% and 19\% of researchers, respectively~\cite{Stuart2018}.

A less-preferred way is an institutional webpage, because some universities do not preserve it when the author changes workplaces. An even worse option is an author's personal homepage, because it is not guaranteed to be maintained. Similar problems may arise with personal (non-institutional) GitHub repositories. More often than not, our experience during the survey was that these personal links did not stand the test of time. In one rather curious case, we discovered (with the help of the Internet Archive / Wayback Machine~\cite{Wayback_machine}) that an author stopped using the domain that hosted a dataset. After the domain name expired and was released to the public market, it was purchased by an unrelated company, which used it to promote online gambling.

\subsubsection{Transparently License the Data for Open Use}
\Cref{sec:intro} discussed that licensing is the second largest \say{pain point} for researchers (37\%)~\cite{Stuart2018}. Although difficulties may arise from the commercial use of data~\cite{rajbahadur2022} or the release of government data~\cite{khayyat2015open}, licensing can be straightforward for many types of LA research. The recommended licenses are summarized by \citet{open-data-licenses}. They include well-known Creative Commons licenses (CC-BY is recommended~\cite{open-data-licenses-nature}, as it is very open but still requires attribution), as well as a dedicated license tailored to open data. The templated definitions of these licenses can be reused directly with minimal or no modifications.
\begin{itemize}
    \item[$\checkmark$] Accompany the dataset with a standard license that allows broad, open use. If possible, we recommend allowing commercial use, as it may enable startups, university spin-offs, and other entities to use the data. Remixing the dataset should also be allowed as long as the original dataset is still cited. Redistribution may be problematic, as the attribution to the original authors can get lost unless the redistributor preserves credit to them.
    \item[$\checkmark$] Ask the users to report to you if they observe the breach of license terms. Such a clause is standard in LA competitions and in prominent datasets, such as those from the ASSISTments ecosystem (see \Cref{tab:datasets}).
\end{itemize}

\subsubsection{Indicate Open Data Availability in Your Paper}
In many articles we examined, it was relatively obvious when a dataset was shared. In the best case, there was an explicit statement such as \say{Data for this article are available at\ldots} or \say{We share the supplementary materials for this paper at\ldots}, accompanied by a clearly visible URL or a reference located in a prominent place, such as the abstract, end of the introduction section, methods, or conclusion (ideally under a subsection such as \say{Resources} or \say{Supplementary Materials}, possibly with a subheading \say{Data}). We implement this practice in \Cref{subsec:implications-materials} and recommend a similar practice also for future authors.
\begin{itemize}
    \item[$\checkmark$] Place the reference to an open dataset in a prominent section of your article.
\end{itemize}

What especially helped was that the template for AIED papers displays all hyperlinks in blue color, which made them immediately stand out during our manual examination. However, most published LAK articles were only black-and-white, even though the \texttt{acmart} template allows the \texttt{screen} option to display colored hyperlinks, which would have helped us. Similarly, for EDM, even though adding the \texttt{hyperref} package to the template to display colored links is allowed, few authors used this option. This could have contributed to accidental false negatives, that is, cases in which we may have missed a dataset that was available. We would like to ask LAK and EDM conference organizers to update their default article templates to incorporate colored hyperlinks, as AIED does. Authors are also advised to use them.
\begin{itemize}
    \item[$\checkmark$] In the PDF version of your paper, display the hyperlinks in color to call visual attention to them.
\end{itemize}

In some cases, the dataset-sharing information was \say{buried} in the article and required a more detailed reading to notice. For example, in datasets stored in Datashop, EDM papers sometimes listed only their numerical ID and no further description, which was a bit obscure. In the worst cases, there is again the risk of false negatives when the description was more hidden, since we did not have time to thoroughly read every full text. Although we tried to mitigate this risk as much as possible, a more explicit pointer from the authors of some articles would make the datasets easier to find.

\subsubsection{Citing Your Dataset Should be Easy for Users}
Most researchers feel they do not receive sufficient credit when others use their datasets~\cite{Science2023}, as discussed in \Cref{sec:intro}. In addition, computing researchers are primarily motivated to share datasets by the prospect of receiving more citations in the future~\cite{Science2023}. 

However, for some of the \numselecteddatasets\ datasets, it was unclear how their authors would like the dataset cited. The default option is to cite the dataset URL, but if the authors prefer an additional attribution (for example, citing a source paper), they should clearly specify it in the dataset description (such as the repository's Readme file). A functional contact email address should also be included (at least in the associated source paper).

The examples of best practice included a clear subheading (e.g., \say{How to cite}) and both full-text and BibTeX citation records that others can simply copy and paste. We recommend BibTeX because it is the most widely supported file format for consistent citations across platforms. This way, the attribution is both correct in the authors' intended sense and easy for the user to incorporate into their article. When compiling \Cref{tab:datasets}, we tried to attribute each dataset to the best of our ability, but in many cases, a guideline from the authors would have helped ensure our citation was correct.
\begin{itemize}
    \item[$\checkmark$] Include the \say{How to cite} section in the description of your dataset to ensure correct attribution by others.
\end{itemize}

\subsubsection{Ensure Responsible Usage and Citation of External Datasets}
In line with the above concern, we also observed several instances in which datasets were not precisely attributed. For example, datasets on PSLC DataShop include fields titled \say{Preferred Citation} and \say{Additional Notes}, where the dataset authors can explain how they would like the dataset attributed. However, even in cases when a dataset had this information listed, we sometimes noticed that when the dataset was used in a secondary paper, the users cited only the PSLC DataShop homepage~\cite{datashop_website}, or the book chapter explaining what PSLC DataShop is~\cite{handbook2010-edm}, but not including the targeted citation of the specific dataset. To avoid this confusion and ensure some degree of standardization in citing datasets, following the dataset authors' instructions is necessary.
\begin{itemize}
    \item[$\checkmark$] Use the \say{How to cite} section in the description of a dataset to ensure you attribute it correctly.
\end{itemize}

\subsection{Supplementary Materials}
\label{subsec:implications-materials}

In the spirit of open science, we provide the following materials for others freely (subject only to the requirement of citing this paper). These resources are available as a DOI-citable release on Zenodo at \url{https://doi.org/10.5281/zenodo.18667402}.
\begin{enumerate}
    \item The research dataset for this survey paper is available in Excel and CSV formats, containing the information we collected about the discovered datasets and papers. All personal identifiers have been removed. The data dictionary explaining each column's meaning is included directly in the spreadsheet.
    \begin{itemize}[topsep=2pt]
        \item[$\circ$] Associated part of the paper: \Cref{sec:method}.
    \end{itemize}
    \item We provide a reproducibility package that includes the Python source code, a requirements file, and execution scripts (\texttt{.sh} for macOS/Linux and \texttt{.bat} for Windows). It builds a containerized environment that automatically analyzes our dataset and generates the results reported in \Cref{sec:results}. The code requires only standard modules, enabling other researchers to extend it and query our dataset for their own insights beyond those in this paper.
    \begin{itemize}[topsep=2pt]
        \item[$\circ$] Associated part of the paper: \Cref{sec:results}.
    \end{itemize}
    \item Email templates used to contact the authors when requesting datasets are available as a PDF.
    \begin{itemize}[topsep=2pt]
        \item[$\circ$] Associated part of the paper: \Cref{subsec:methods-criteria}.
    \end{itemize}
    \item We provide BibTeX files in a unified, pretty-print format that have been triple-checked by independent validators. These files support the correct citation of the \numselecteddatasets\ datasets and \numselectedpapers\ papers if others decide to pursue them further.
    \begin{itemize}[topsep=2pt]
        \item[$\circ$] Associated part of the paper: \Cref{sec:appendix-datasets}.
    \end{itemize}
\end{enumerate}

\begin{acks}
\textit{(1) Funding acknowledgments:}
\begin{itemize}
    \item The work of Valdemar Švábenský on this paper was supported by the Czech Science Foundation (GAČR) grant no. 25-15839I.
    \item This work was supported by JST CREST Grant Number JPMJCR22D1 and JSPS KAKENHI Grant Number JP22H00551, Japan.
    \item This work was partly supported by JSPS Grant-in-Aid for Scientific Research (B) JP23H01001 and JP22H03902, (Exploratory) JP21K19824, (International-B) JP24KK0051.
\end{itemize}
\textit{(2) Thanks:}
\begin{itemize}
    \item We thank Ryan S. Baker for providing useful ideas regarding the paper's scope and research questions during brainstorming sessions at an early stage of writing.
    \item We also thank Daiki Matsumoto, Ryusei Munemura, and Sota Totoki for preparing the BibTeX citation records for all selected papers and datasets, as well as validating and correcting the final dataset table.
    \item Finally, we thank the anonymous reviewers for providing helpful suggestions that substantially improved the paper during the major revision phase.
\end{itemize}
\textit{(3) Dataset usage (all listed datasets are cited in \Cref{tab:datasets}, but some of them required including acknowledgment texts in addition to their citations):}
\begin{itemize}
    \item We would like to thank Carnegie Learning, Inc., for providing the Cognitive Tutor data supporting this analysis.
    \item This research used the \say{RIKEN Essay Question Scoring Dataset,} provided by the National Institute of Informatics' IDR Dataset Provision Service, RIKEN, a national research and development agency.
\end{itemize}

\end{acks}

\bibliographystyle{ACM-Reference-Format}
\bibliography{references-main, references-selected-datasets, references-selected-papers}

\appendix

\newpage

\newcommand{\popular}{\scalebox{0.75}{$\star$} }

{
\renewcommand{\arraystretch}{1.025}
\footnotesize
\rowcolors{2}{gray!15}{white}

\begin{longtable}{rll}

\caption{All \numselecteddatasets\ datasets sorted alphabetically. The \popular marks datasets used 3+ times. URL validity last checked on 31 January 2025.}
\label{tab:datasets}
\\  

\toprule
\rowcolor{white}
\textbf{ID} & \textbf{Dataset name as URL, description, and citation} & \textbf{Usage in papers at LAK, EDM, and AIED} \\ 
\midrule
\endfirsthead

\rowcolor{white}
\caption*{\Cref{tab:datasets} continued.}
\\

\toprule
\rowcolor{white}
\textbf{ID} & \textbf{Dataset name as URL, description, and citation} & \textbf{Usage in papers at LAK, EDM, and AIED} \\ 
\midrule
\endhead

\bottomrule
\endfoot


1 & \href{https://www.kaggle.com/datasets/septa97/100k-courseras-course-reviews-dataset}{100K Coursera's Course Reviews} -- labeled user reviews of online courses \cite{100K-Coursera-Course-Reviews-Dataset} & 1\texttimes~\cite{fAIED22fine-grained} \\

2 & \href{https://sharedtask.duolingo.com/2018.html}{2018 Duolingo Second Language Acquisition Modeling (SLAM) Task} \cite{2018-Duolingo-Shared-Task, 2018-Duolingo-Shared-Task2} & 2\texttimes~\cite{fLAK21variational, fEDM20variational} \\

3 & \href{https://data.mendeley.com/datasets/83tcx8psxv/1}{Academic Performance Evolution for Engineering Students} -- assessment \cite{AcadPerfEvolEng} & 1\texttimes~\cite{fEDM23investigating} \\

4 & \href{https://gitlab.com/JonathanNau/argument-annotated-essays}{Argument Annotated Essays (Nau)} -- sentences from Portuguese essays  \cite{Argument-Annotated-Essays} & 1\texttimes~\cite{fLAK22towards} \\

\popular 5 & \href{https://www.kaggle.com/c/asap-aes/data}{ASAP-AES (Automated Essay Scoring Prize)} -- essays and their grades \cite{asap-aes} & 7\texttimes~\cite{sEDM21integrating, sEDM21on, fEDM22individual, fEDM23evaluating, sAIED21assessment2vec, sAIED21integration, fAIED23neural} \\

\popular 6 & \href{https://www.kaggle.com/competitions/asap-sas}{ASAP-SAS (Short Answer Scoring Prize)} -- textual answers to questions \cite{asap-sas} & 4\texttimes~\cite{fAIED22balancing-cost, fAIED22representing, fAIED23generalizable, fAIED24explainable} \\

7 & \href{https://github.com/Aries-chen/MTAA/tree/main/dataset/ASAP%2B%2B}{ASAP++} -- annotated version of ASAP-AES enhanced with attribute scores \cite{asap++} & 2\texttimes~\cite{fLAK23towards, sAIED24a} \\

8 & \href{https://drive.google.com/drive/folders/18D5iXB0GtAEM4Zu5JhoHudnSQu5b1eQ2}{ASSISTments: Affect Original} -- student interaction logs \& affect labels \cite{ASSIST, ASSISTments-affect-original2} & 1\texttimes~\cite{fAIED22deep} \\

9 & \href{https://drive.google.com/drive/folders/1b-vxAh5hEqbbvIJNLn6aIBmWQ-Pd6IvL}{ASSISTments: Affect with Urbanicity} -- logs with additional labels \cite{ASSIST, ASSISTments-affect-with-urbanicity2} & 2\texttimes~\cite{fLAK21using, sLAK21using} \\

10 & \href{https://osf.io/sfyzv/?view_only=351fb8781d2c4f3bbc9d7486762d563a}{ASSISTments: Bandit Simulation} -- for evaluation of algorithms \cite{ASSIST, sLAK24expert} & 1\texttimes~\cite{sLAK24expert} \\

11 & \href{https://github.com/AshishJumbo/LAK_CWAF}{ASSISTments: Common Wrong Answers} -- student activity data \cite{ASSIST, fLAK23identification} & 1\texttimes~\cite{fLAK23identification} \\

12 & \href{https://osf.io/59shv/}{ASSISTments: E-TRIALS} -- combined data from 50 experiments \cite{ASSIST, fEDM22exploring} & 1\texttimes~\cite{fEDM22exploring} \\

13 & \href{https://github.com/kirkvanacore/Non-Cognitive-Interventions_ASSSITments/tree/main/Data}{ASSISTments: Non-cognitive} -- students' work on problem sets \cite{ASSIST, fLAK23impact} & 1\texttimes~\cite{fLAK23impact} \\

14 & \href{https://osf.io/spg6v/}{ASSISTments: Reaction to Anticipation} -- student interactions \& affect \cite{ASSIST, sEDM24from} & 1\texttimes~\cite{sEDM24from} \\

15 & \href{https://www.openicpsr.org/openicpsr/project/183645/version/V1/view}{ASSISTments: Replication Study} -- student and teacher data \cite{ASSIST, fAIED23implementing} & 1\texttimes~\cite{fAIED23implementing} \\

16 & \href{https://drive.google.com/drive/folders/1fRhyVEetIsgRdp-B8J5seH64FCHC2HMI}{ASSISTments: Response Time Decomposition} -- interaction logs \cite{ASSIST, fLAK21examining} & 1\texttimes~\cite{fLAK21examining} \\

17 & \href{https://osf.io/uj48v/}{ASSISTments: Surrogate Measures} -- students' clickstream sequences \cite{ASSIST, sEDM23effective} & 1\texttimes~\cite{sEDM23effective} \\

\popular 18 & \href{https://sites.google.com/site/assistmentsdata/home/2009-2010-assistment-data}{ASSISTments2009-10} -- math skill builder dataset, removed duplicates \cite{ASSIST, ASSISTments2009-2} & 12\texttimes~\cite{sLAK24architectural, sEDM20deep, sEDM20dynamic, fEDM21student, sEDM21deep-irt, sEDM21do, sEDM21pybkt, sAIED20learning, sAIED20towards, sAIED21pakt, fAIED22deep, sAIED22a} \\

\popular 19 & \href{https://sites.google.com/site/assistmentsdata/datasets/2012-13-school-data-with-affect}{ASSISTments2012-13-affect} -- knowledge acquisition \& learner affect \cite{ASSIST, ASSISTments2012-affect2} & 7\texttimes~\cite{fLAK21learning, sEDM20dynamic, fEDM21student, fEDM22automatic, sEDM23auto-scoring, sEDM23modeling, sAIED20predicting} \\

\popular 20 & \href{https://sites.google.com/site/assistmentsdata/datasets/2015-assistments-skill-builder-data}{ASSISTments2015-16} -- newer skill builder dataset with problem sets \cite{ASSIST, ASSISTments2015-2} & 4\texttimes~\cite{sEDM21deep-irt, fAIED20early, sAIED20learning, sAIED21pakt} \\

21 & \href{https://sites.google.com/site/las2016data/data/thison?authuser=0}{ASSISTments2016-LAS} -- combined data from 22 experiments \cite{ASSIST, ASSISTments2016-LAS2} & 1\texttimes~\cite{fEDM20getting} \\

\popular 22 & \href{https://sites.google.com/view/assistmentsdatamining/dataset}{ASSISTments2017 (ASSISTChal)} -- longitudinal student data \cite{ASSIST, ASSISTments2017-2} & 9\texttimes~\cite{fLAK24long-term, sLAK24architectural, sEDM20deep, fEDM21student, sAIED20deep, fAIED21deep, fAIED21seven-year, sAIED21incorporating, sAIED22self-attention} \\

23 & \href{https://sites.google.com/view/dataminingcompetition2019/dataset?authuser=0}{ASSISTments2019-NAEP (Nation's Report Card DM Competition)} \cite{ASSIST} & 1\texttimes~\cite{fEDM20the-NAEP} \\

24 & \href{https://github.com/readerbench/AMoC}{Automated Model of Comprehension (AMoC)} -- texts from various domains \cite{fAIED23the} & 1\texttimes~\cite{fAIED23the} \\

25 & \href{https://github.com/AntonetteShibani/AutomatedRevisionGraphs}{Automated Revision Graphs (ARG)} -- student revisions of their writing \cite{sAIED20constructing} & 1\texttimes~\cite{sAIED20constructing} \\

26 & \href{https://github.com/epfl-ml4ed/beerslaw-lab}{Beer's Law} -- students' clickstream data and responses to tasks \cite{fEDM22generalisable} & 1\texttimes~\cite{fEDM22generalisable} \\
        
27 & \href{https://eric.ed.gov/?id=ED054233}{Bormuth Corpus} -- English text passages (from a 1971 study of readability) \cite{Bormuth-Corpus} & 1\texttimes~\cite{fAIED22assessing} \\

28 & \href{https://www.gov.br/inep/pt-br/acesso-a-informacao/dados-abertos/microdados/enem}{Brazilian National High School Exam (ENEM Microdados)} \cite{Brazilian-National-High-School-Exam} & 1\texttimes~\cite{fLAK23predicting} \\

29 & \href{https://osf.io/vgrfk/?view_only=8579adb171a64c07bb6641be67dba202}{Bridging Learnersourcing and AI} -- written hints to programming tasks \cite{sLAK24bridging} & 1\texttimes~\cite{sLAK24bridging} \\

30 & \href{https://dataverse.harvard.edu/dataset.xhtml?persistentId=doi:10.7910/DVN/1XORAL}{Canvas Network (CanvasNet)} -- students' interactions with MOOCs \cite{CanvasNet} & 2\texttimes~\cite{fEDM20analyzing, fEDM20modeling} \\

31 & \href{https://github.com/yixin-cheng/coAuthor/tree/main/coauthor-v1.0}{CoAuthor} -- writing sessions with generative AI (GenAI) \cite{CoAuthor} & 2\texttimes~\cite{fLAK24evidence-centered, fEDM23visual} \\

32 & \href{https://github.com/DSAatUSU/ChatGPT-promises-and-pitfalls}{Code Generation Prompts With Code Samples} -- for GenAI applications \cite{fEDM24assessing} & 1\texttimes~\cite{fEDM24assessing} \\

\popular 33 & \href{https://pslcdatashop.web.cmu.edu/DatasetInfo?datasetId=613}{CogTutor (Algebra I, CTA)} -- students' responses to math questions \cite{datashop, CogTutor2} & 5\texttimes~\cite{fEDM20feature, fEDM20the-ebb, fEDM20the-effect, sEDM22evaluating, fAIED23algebra} \\

34 & \href{https://pslcdatashop.web.cmu.edu/DatasetInfo?datasetId=392}{CogTutor (Geometry)} -- experiment data of learning improvement \cite{datashop, CogTutor2, CogTutorSimGeomArea2} & 2\texttimes~\cite{sEDM21toward, sEDM24replicating} \\

35 & \href{https://pslcdatashop.web.cmu.edu/DatasetInfo?datasetId=99}{CogTutor (High School Geometry)} -- Carnegie Learning and CTAT \cite{datashop, CogTutor2} & 1\texttimes~\cite{sEDM24replicating} \\
        
36 & \href{https://pslcdatashop.web.cmu.edu/DatasetInfo?datasetId=253}{CogTutor (High School Geometry Area)} -- similar as above \cite{datashop, CogTutor2} & 1\texttimes~\cite{sEDM24replicating} \\
        
37 & \href{https://pslcdatashop.web.cmu.edu/DatasetInfo?datasetId=2174}{CogTutor (Simulated -- Geometry Area)} -- synthetic transactions \cite{datashop, CogTutor2, CogTutorSimGeomArea2} & 1\texttimes~\cite{sEDM20confident} \\

38 & \href{https://pslcdatashop.web.cmu.edu/DatasetInfo?datasetId=1007}{College Computer Science (Mathan)} -- test and interaction data \cite{datashop, CollegeCompSci2} & 1\texttimes~\cite{sEDM24replicating} \\

39 & \href{https://pslcdatashop.web.cmu.edu/DatasetInfo?datasetId=104}{College Physics (Self-explanation)} -- in vivo experiment data \cite{datashop, CollegePhysics2} & 1\texttimes~\cite{sEDM24replicating} \\

40 & \href{https://wilburone.github.io/cosmos/}{CosmosQA} -- reading comprehension multiple-choice questions (MCQs) \cite{CosmosQA} & 1\texttimes~\cite{fEDM24disto} \\

41 & \href{https://pslcdatashop.web.cmu.edu/Files?datasetId=2865}{CS EDM 2019 Challenge} -- novice programmers' data from an ITS \cite{datashop} & 1\texttimes~\cite{sAIED22self-attention} \\

\popular 42 & \href{https://pslcdatashop.web.cmu.edu/DatasetInfo?datasetId=3458}{CS EDM 2020 Challenge (CodeWorkout)} -- students' Java code \cite{datashop, CSEDM-2020-2} & 4\texttimes~\cite{sEDM21more, fEDM23analysis, fEDM23kc-finder, sEDM24evaluating} \\

43 & \href{https://drive.google.com/drive/folders/1NwG4H071MtDvjlmfzFuceeoHLAA0cXsf}{DECEP: Educational Content Effectiveness using Physiological data} \cite{fLAK24effecti-net} & 1\texttimes~\cite{fLAK24effecti-net} \\

44 & \href{https://github.com/joyheyueya/declarative-math-word-problem}{Declarative Math Word Problems} -- tasks from algebra textbooks \cite{Declarative-Math-Word-Problems} & 1\texttimes~\cite{fEDM24evaluating} \\

45 & \href{https://bitbucket.org/iiscseal/deepfix/src/master/}{DeepFix (Prutor)} -- C programs that fail to compile \cite{DeepFix} & 2\texttimes~\cite{fAIED20macer, fAIED21repairnet} \\

46 & \href{https://github.com/deepmind/mathematics_dataset}{DeepMind Maths} -- school-level math question and answer pairs \cite{DeepMindMaths} & 1\texttimes~\cite{sEDM21math} \\

47 & \href{https://github.com/DRSY/DGen/tree/main/Layer1/dataset}{Distractor Generation (DGen)} -- fill-in-the-blank questions \cite{Distractor-Generation} & 1\texttimes~\cite{sAIED24beyond} \\

48 & \href{https://dataset.org/dream/}{DREAM: Dialogue-based REAding comprehension exaMination} -- MCQs \cite{DREAM} & 1\texttimes~\cite{fEDM24disto} \\

\popular 49 & \href{https://github.com/duolingo/halflife-regression}{Duolingo Spaced Repetition: Historical Log dataset (HLD)} -- traces \cite{Duolingo-spaced-repetition, Duolingo-spaced-repetition-2} & 4\texttimes~\cite{fLAK21linguistic, fLAK21variational, fLAK23each, sAIED20adaptive} \\

\popular 50 & \href{https://github.com/riiid/ednet}{EdNet by Riiid} -- student interactions with an English tutoring service \cite{sAIED20ednet} & 11\texttimes~\cite{fLAK21recommendation, sLAK21saint, fEDM21learning-from, sEDM21knowledge, sEDM21lana, fEDM22optimizing, sAIED20ednet, fAIED21option, sAIED23multi-dimensional, fAIED24federated, fAIED24knowledge} \\

\pagebreak

51 & \href{https://github.com/hadifar/question-generation}{EduQG} -- multiple-choice questions (MCQs) for question generation \cite{EduQG} & 1\texttimes~\cite{sAIED24towards} \\

52 & \href{https://hmi.iiitd.edu.in/engagemeDataset.html}{EngageME} -- measures of student engagement to assess attention \cite{fAIED24engageme} & 1\texttimes~\cite{fAIED24engageme} \\

53 & \href{https://pslcdatashop.web.cmu.edu/DatasetInfo?datasetId=313}{English Articles 2009 (IWT Self-Explanation Study 1)} \cite{datashop, EngArticles2} & 1\texttimes~\cite{sEDM21toward} \\

54 & \href{https://pslcdatashop.web.cmu.edu/DatasetInfo?datasetId=372}{English Articles 2009-2 (IWT Self-Explanation Study 2)} \cite{datashop, EngArticles2} & 1\texttimes~\cite{sEDM21toward} \\

55 & \href{https://pslcdatashop.web.cmu.edu/DatasetInfo?datasetId=394}{English Articles 2010 (IWT Self-Explanation Study 3)} \cite{datashop, EngArticles2} & 1\texttimes~\cite{sEDM21toward} \\

56 & \href{https://github.com/AIED2021/ESL-SentenceCompletion/tree/main}{ESL (English as a Second Language) Sentence Completion} \cite{sAIED21solving} & 1\texttimes~\cite{sAIED21solving} \\

57 & \href{http://yoehara.com/EVKD1/}{ESL learners' Vocabulary Knowledge Dataset (EVKD1)} \cite{ESL-learners-Vocabulary-Knowledge, ESL-learners-Vocabulary-Knowledge2} & 2\texttimes~\cite{sEDM22no, fAIED22an} \\

58 & \href{https://github.com/rafaelanchieta/essay}{Essay-BR} -- argumentative texts in Brazilian Portuguese \cite{Essay-BR} & 1\texttimes~\cite{fLAK23towards} \\

59 & \href{https://aic-fe.bnu.edu.cn/cgzs/kfsj/xxszskszw/index.html}{Essay-CN: Formal Examination} -- student essays in Chinese \cite{fAIED24leveraging} & 1\texttimes~\cite{fAIED24leveraging} \\

60 & \href{https://github.com/uci-soe/FairytaleQAData}{FairytaleQA} -- questions derived from 278 children-friendly stories \cite{FairytaleQA} & 1\texttimes~\cite{sAIED23towards} \\

61 & \href{https://pslcdatashop.web.cmu.edu/DatasetInfo?datasetId=5549}{Feedback: Dynamic Transitions} -- annotated log data incl. feedback \cite{datashop, fEDM24combining} & 1\texttimes~\cite{fEDM24combining} \\

62 & \href{https://pslcdatashop.web.cmu.edu/DatasetInfo?datasetId=5153}{Feedback: Hopewell Collaboration} -- annotated log data incl. feedback \cite{datashop, fEDM24combining} & 1\texttimes~\cite{fEDM24combining} \\

63 & \href{https://pslcdatashop.web.cmu.edu/DatasetInfo?datasetId=5604}{Feedback: Steel Valley} -- annotated log data incl. feedback \cite{datashop, fEDM24combining} & 1\texttimes~\cite{fEDM24combining} \\

64 & \href{https://www.kaggle.com/competitions/feedback-prize-english-language-learning}{Feedback Prize: ELLIPSE (English Language Learners)} -- student essays \cite{Feedback-Prize-ELLIPSE} & 1\texttimes~\cite{sAIED24a} \\

65 & \href{https://www.kaggle.com/competitions/feedback-prize-2021}{Feedback Prize: ESW (Evaluating Student Writing)} -- student essays \cite{Feedback-Prize-ESW} & 1\texttimes~\cite{fAIED23“Why} \\

66 & \href{https://osf.io/t8wzy/}{Figures Similarity} -- participants' ratings of similarity of images \cite{Figures-Similarity} & 1\texttimes~\cite{sEDM24integrating} \\

67 & \href{https://github.com/delmarin35/Dynamic-Neural-Models-for-Knowledge-Tracing/tree/main/Datasets/fsaif1tof3}{Find Solution AI Limited (FSAI-F1toF3)} -- student data from a virtual TA \cite{fEDM21student} & 1\texttimes~\cite{fEDM21student} \\

68 & \href{http://staff.ustc.edu.cn/~qiliuql/data/math2015.rar}{FracSub (Math2015)} -- graded student responses to math problems \cite{FracSub} & 1\texttimes~\cite{fLAK20applying} \\

69 & \href{https://pslcdatashop.web.cmu.edu/DatasetInfo?datasetId=1190}{Fraction Addition and Multiplication (Blocked vs. Interleaved)} \cite{datashop, fAIED20investigating} & 1\texttimes~\cite{fAIED20investigating} \\

70 & \href{https://pslcdatashop.web.cmu.edu/DatasetInfo?datasetId=671}{Fractions Lab Experiment} -- student interactions with an ITS \cite{datashop, Fractions-Lab-Experiment} & 1\texttimes~\cite{fAIED24beyond} \\

71 & \href{https://osf.io/ehm43/}{Gamified Review Assessments} -- students' self-reported motivation \cite{sAIED22garfield} & 1\texttimes~\cite{sAIED22garfield} \\

72 & \href{https://github.com/Mocahteam/GdPMOOC_Fairness}{Gestion de Projet (GdP MOOC)} -- course logs, grades, and demographics \cite{fAIED24fairness} & 1\texttimes~\cite{fAIED24fairness} \\

73 & \href{https://www.oecd.org/en/about/projects/global-teaching-insights.html}{Global Teaching Insights (GTI)} -- video study of instructional practices \cite{Global-Teaching-Insights} & 1\texttimes~\cite{fAIED24automated-assessment} \\

74 & \href{https://github.com/google-research/google-research/tree/master/goemotions}{GoEmotions} -- Reddit comments manually labeled with emotion categories \cite{GoEmotions} & 1\texttimes~\cite{fLAK24predicting} \\

75 & \href{https://dataverse.harvard.edu/dataset.xhtml?persistentId=doi:10.7910/DVN/26147}{HarvardX and MITx} -- student activities and grades from MOOCs \cite{HarvardX-and-MITx} & 1\texttimes~\cite{sEDM21fair} \\

76 & \href{https://nces.ed.gov/datalab/onlinecodebook/session/codebook/70e46c64-4233-44a2-b775-135468d0784a}{High School Longitudinal Study (HSLS)} -- data from various stakeholders \cite{HighSchoolLongitudinalStudy} & 1\texttimes~\cite{sEDM24examining} \\

77 & \href{https://github.com/IEClab-NCSU/SimStudent/tree/main/CTI}{Human Tutor Responses: Constructive Tutee Inquiry} -- labeled responses \cite{fAIED23what} & 1\texttimes~\cite{fAIED23what} \\

78 & \href{https://osf.io/daejr/}{Individual Temporal Behavior} -- students’ trace data within a course \cite{Individual-Temporal-Behavior-1, fAIED24from} & 1\texttimes~\cite{fAIED24from} \\

79 & \href{https://github.com/chili-epfl/induce/blob/master/analysis/data/induce-data-2019-08-08.csv}{Induce} -- student answers to logic quiz questions on inductive reasoning \cite{fLAK20a} & 1\texttimes~\cite{fLAK20a} \\

80 & \href{https://huggingface.co/datasets/koutch/intro_prog}{Intro Programming: Dublin and Singapore} -- student submissions \cite{intro_prog-1, sAIED23automated} & 2\texttimes~\cite{sAIED23automated, sAIED23training} \\

81 & \href{https://github.com/GCleuziou/code2aes2vec}{Intro Programming: Dublin-42487} -- student programs enriched by tests \cite{fEDM21learning-student, intro_prog-Dublin-42487-2} & 1\texttimes~\cite{fEDM21learning-student} \\

82 & \href{https://github.com/GCleuziou/code2aes2vec}{Intro Programming: NewCaledonia-5690} -- short Python programs \cite{fEDM21learning-student, intro_prog-Dublin-42487-2} & 1\texttimes~\cite{fEDM21learning-student} \\

83 & \href{https://pslcdatashop.web.cmu.edu/DatasetInfo?datasetId=1198}{Junyi Academy Math Practicing Log} -- student problem-solving \cite{datashop, Junyi-2} & 1\texttimes~\cite{fEDM21learning-from} \\

84 & \href{https://zenodo.org/records/4627104}{JUSThink Dialogue and Actions Corpus} -- transcripts and logs \cite{JUSThink-Dialogue-and-Actions-Corpus-1, JUSThink-Dialogue-and-Actions-Corpus-2} & 1\texttimes~\cite{fEDM23to} \\

85 & \href{https://zenodo.org/records/13834073}{JUSThink: PE-HRI-temporal} -- team behaviors and learning outcomes \cite{JUSThink-Dialogue-and-Actions-Corpus-1} & 1\texttimes~\cite{fEDM23to} \\

\popular 86 & \href{https://pslcdatashop.web.cmu.edu/KDDCup/downloads.jsp}{KDD Cup 2010 (Challenge and Development)} -- five math datasets \cite{datashop, KDD-Cup-2010, KDD-Cup-2010-2} & 7\texttimes~\cite{fLAK21learning, sLAK21modelling, sEDM21deep-irt, sEDM21do, fEDM23scalable, sAIED20deep, sAIED21individualization} \\

87 & \href{https://github.com/CAHLR/skill-equivalency/tree/main/data}{Khan Academy} -- student responses to math exercises \cite{fLAK21learning} & 1\texttimes~\cite{fLAK21learning} \\

88 & \href{https://github.com/SteveLEEEEE/EDM2022CLO}{Learning Objectives of 5,558 Courses} -- at an Australian university \cite{sEDM22automatic} & 1\texttimes~\cite{sEDM22automatic} \\

89 & \href{https://huggingface.co/datasets/sidovic/LearningQ-qg}{LearningQ-qg} -- dataset for educational question generation (QG) \cite{LearningQ-qg} & 1\texttimes~\cite{fAIED20remember} \\

90 & \href{https://github.com/jvasso/graph-rl4adaptive-learning}{Linear Corpus of Course Descriptions} -- topic descriptions from courses \cite{sEDM23towards-scalable} & 1\texttimes~\cite{sEDM23towards-scalable} \\

91 & \href{https://osf.io/ksmqe/}{LLM-Based Chatbots' Written Prompts} -- assessing cultural intelligence \cite{fAIED24on} & 1\texttimes~\cite{fAIED24on} \\

92 & \href{https://osf.io/u4ntf/}{Lost in Translation} -- LMS activity records across multiple courses \cite{fLAK23lost} & 1\texttimes~\cite{fLAK23lost} \\

93 & \href{https://zenodo.org/records/1487858}{MCQs from MOOCs} -- questions in 18 MOOCs assessed for flaws \cite{MCQs-from-MOOCs} & 1\texttimes~\cite{fAIED24an} \\

94 & \href{https://my.hidrive.com/share/wdnind8pp5#$/}{MCScript2.0 (Machine Comprehension)} -- crowdsourced questions \cite{MCScript2.0} & 1\texttimes~\cite{fEDM24disto} \\

95 & \href{https://github.com/JacksonWuxs/MeNSP}{MeNSP (Next Sentence Prediction)} -- written responses to questions \cite{fAIED23matching, MeNSP-2} & 1\texttimes~\cite{fAIED23matching} \\

96 & \href{https://osf.io/mej2r/?view_only=10f6bbfb01084a2698ec326ed1a1c03d}{Millions of Views} -- records of learners who watched educational videos \cite{fLAK24millions} & 1\texttimes~\cite{fLAK24millions} \\

97 & \href{https://github.com/UM-Lifelong-Learning-Lab/AIED2022-MixFA-dataset}{MixFA (Formative Assessments)} -- questions with student answers \cite{fAIED22scaling} & 1\texttimes~\cite{fAIED22scaling} \\

\popular 98 & \href{http://infolab.stanford.edu/~paepcke/stanfordMOOCForumPostsSet.tar.gz}{MOOC Forum Posts: Stanford} -- learner posts from online classes \cite{MOOC-Forum-Posts-Stanford} & 4\texttimes~\cite{sEDM20effective, fEDM21which, sEDM23towards-generalizable, sAIED20bert} \\

99 & \href{https://github.com/pcla-code/forum-posts-urgency}{MOOC Forum Posts: UPenn} -- learner posts from university courses \cite{sEDM23towards-generalizable} & 1\texttimes~\cite{sEDM23towards-generalizable} \\

100 & \href{https://www.frontiersin.org/journals/education/articles/10.3389/feduc.2022.736194/full}{Multimodal Analytics} -- features based on audio/video recordings \cite{Multimodal-analytics-1, fLAK23how} & 2\texttimes~\cite{fLAK23how, sLAK23impact} \\

\pagebreak

101 & \href{https://myDALITE.org/signup}{myDALITE} -- learnersourced explanations from an online environment \cite{sEDM20a} & 1\texttimes~\cite{sEDM20a} \\

102 & \href{https://github.com/ddemszky/conversational-uptake}{NCTE (Center for Teacher Effectiveness) Uptake} -- student-teacher dialogue \cite{NCTE-uptake} & 1\texttimes~\cite{sEDM22the} \\

\popular 103 & \href{https://eedi.com/projects/neurips-education-challenge}{NeurIPS 2020 Challenge} -- options students selected in MCQs \cite{NeurIPS-2020-1, NeurIPS-2020-2} & 6\texttimes~\cite{fLAK24improving, sEDM21quizzing, fEDM22sparse, fAIED21deep, fAIED21option, sAIED24using} \\

104 & \href{https://www.kaggle.com/datasets/nltkdata/nps-chat}{NPS (Naval Postgraduate School) Chat Corpus} -- labeled chat posts \cite{NPS-Chat-Corpus} & 1\texttimes~\cite{fLAK24predicting} \\

\popular 105 & \href{https://pslcdatashop.web.cmu.edu/DatasetInfo?datasetId=507}{OLI Engineering Statics} -- student skills in an engineering course \cite{datashop} & 6\texttimes~\cite{sEDM21deep-irt, sAIED20deep, sAIED20learning, fAIED21deep, sAIED21incorporating, sAIED22self-attention} \\

106 & \href{https://zenodo.org/records/1219041}{OneStopEnglish Corpus} -- for assessing readability and simplifying texts \cite{OneStopEnglish-Corpus} & 1\texttimes~\cite{fAIED22assessing} \\

107 & \href{https://geiser.github.io/phd-thesis-evaluation/study03/data/}{Ontology-based Gamified Sessions} -- student responses in a questionnaire \cite{fAIED20can} & 1\texttimes~\cite{fAIED20can} \\

108 & \href{https://github.com/openstax/research-question-generation-gpt3}{OpenStax Biology} -- sources for LLM question generation \cite{fAIED22towards-human-like} & 1\texttimes~\cite{fAIED22towards-human-like} \\

\popular 109 & \href{https://analyse.kmi.open.ac.uk/open-dataset}{OULAD (Open University LA Dataset)} -- course data and demographics \cite{OULAD} & 6\texttimes~\cite{sEDM20online, sEDM21fair, fEDM23investigating, fEDM23is, fAIED23trustworthy, fAIED24federated} \\

110 & \href{https://paracrawl.eu/}{ParaCrawl} -- corpora of equivalent documents in various languages \cite{ParaCrawl} & 1\texttimes~\cite{fAIED21multilingual} \\

111 & \href{http://www.tml.cs.uni-tuebingen.de/team/luxburg/code_and_data/peer_grading_data_request.php}{Peer Grading in a Computer Science Course} -- grading data \cite{PeerGrading} & 1\texttimes~\cite{sEDM22improving} \\

112 & \href{https://tudatalib.ulb.tu-darmstadt.de/handle/tudatalib/2422}{Persuasive Essays} -- texts with annotations of argument components \cite{Persuasive-Essays} & 1\texttimes~\cite{sEDM20incorporating} \\

113 & \href{https://github.com/AmirZur/smartstem-ai}{Physics and Chemistry Questions} -- questions and learning objectives \cite{sEDM23meta-learning} & 1\texttimes~\cite{sEDM23meta-learning} \\

114 & \href{https://github.com/luffycodes/Tutorbot-Spock-Phys/}{Physics Questions (PHY300)} -- questions and LLM conversations \cite{sLAK24code} & 1\texttimes~\cite{sLAK24code} \\

115 & \href{https://github.com/pkt-attn/pkt-attn}{PKT-ATTN (Programming Knowledge Tracing + Attention)} -- exercise logs \cite{sLAK22enhancing} & 1\texttimes~\cite{sLAK22enhancing} \\

116 & \href{https://www.microsoft.com/en-us/download/details.aspx?id=52397}{Powergrading: Short Answer Grading Corpus} -- US citizenship test \cite{PowerGrading} & 1\texttimes~\cite{fEDM21generative} \\

117 & \href{https://stanford.edu/~cpiech/pyramidsnapshot/challenge.html}{PyramidSnapshot} -- annotated images from programming task snapshots \cite{PyramidSnapshot} & 1\texttimes~\cite{fEDM21generative} \\

118 & \href{https://www.cs.cmu.edu/~ark/QA-data/}{QA (Question-Answer) Dataset} -- Wikipedia articles and questions \cite{QA-data} & 1\texttimes~\cite{fAIED22auxiliary} \\

119 & \href{https://github.com/VenkteshV/QDIFF_AIED_2022}{QC-Science} -- question-answer pairs tagged with Bloom’s taxonomy \cite{fAIED22auxiliary} & 1\texttimes~\cite{fAIED22auxiliary} \\

120 & \href{https://text-machine.cs.uml.edu/lab2/projects/quail/}{QuAIL (Question Answering for AI)} -- question-answer dataset \cite{QuAIL} & 1\texttimes~\cite{fEDM24disto} \\

121 & \href{https://github.com/jkoppel/QuixBugs}{QuixBugs} -- programs with 1-line defects, along with test cases \cite{QuixBugs} & 1\texttimes~\cite{sAIED23automated} \\

122 & \href{https://www.cs.cmu.edu/~glai1/data/race/}{RACE (ReAding Comprehension Dataset From Examinations)} \cite{RACEq} & 2\texttimes~\cite{fEDM24disto, fAIED20remember} \\

123 & \href{https://github.com/DigitalHarborFoundation/rag-for-math-qa}{RAG (Retrieval-Augmented Generation) for Math Question Answering} \cite{sEDM24retrieval-augmented} & 1\texttimes~\cite{sEDM24retrieval-augmented} \\

124 & \href{https://pslcdatashop.web.cmu.edu/DatasetInfo?datasetId=1195}{REAL Genetics} -- student interactions with a problem-solving tutor \cite{datashop, REAL-Genetics2} & 1\texttimes~\cite{fAIED24beyond} \\

125 & \href{https://engineering.purdue.edu/coursemirror/download/reflections-quality-data/}{Reflections Quality (CourseMIRROR)} -- annotated student reflections \cite{Reflections-Quality, Reflections-Quality2} & 1\texttimes~\cite{sAIED22improving} \\

126 & \href{https://github.com/luffycodes/Automated-Long-Answer-Grading}{RiceChem} -- student answers with their grading by teachers \cite{fAIED24automated-long} & 1\texttimes~\cite{fAIED24automated-long} \\

127 & \href{https://www.kaggle.com/c/riiid-test-answer-prediction/data/}{Riiid} -- question answer correctness prediction on an English test \cite{Riiid} & 2\texttimes~\cite{sEDM21lana, sAIED21an} \\

\popular 128 & \href{https://www.nii.ac.jp/dsc/idr/rdata/RIKEN-SAA/}{RIKEN Essay Questions Scoring} -- answers to Japanese questions \cite{RIKEN-Essay-Questions-Scoring1, RIKEN-Essay-Questions-Scoring2} & 3\texttimes~\cite{fAIED22balancing-cost, fAIED22plausibility, fAIED23reducing} \\

129 & \href{https://drive.google.com/file/d/1oivtasEHGpgRQ9n1WKOtcHkebnvx7xDC/view}{Robomission (Blockly Programming Data)} -- block programming task data \cite{Robomission} & 2\texttimes~\cite{fLAK20exploration, sAIED20automated} \\

130 & \href{https://github.com/sruettgers/automatic_matchmaking}{RoundNet} -- team sport data for automated matchmaking \cite{sEDM24automatic} & 1\texttimes~\cite{sEDM24automatic} \\

131 & \href{https://github.com/allenai/s2orc/}{S2ORC: Semantic Scholar Open Research Corpus} -- academic papers \cite{S2ORC} & 1\texttimes~\cite{fAIED23scalable} \\

132 & \href{https://github.com/jilljenn/qna}{Scholastic Assessment Test (SAT)} -- responses to various questions \cite{Scholastic-Assessment-Test1, Scholastic-Assessment-Test2} & 1\texttimes~\cite{fLAK20applying} \\

\popular 133 & \href{https://allenai.org/data/sciq}{SciQ (Science Questions)} -- crowdsourced exam questions \cite{SciQ} & 3\texttimes~\cite{fEDM24disto, fAIED23scalable, fAIED24fine-tuning} \\

134 & \href{https://osf.io/pyb8s/}{Second Language Acquisition (CritLangAcq)} -- English quiz questions \cite{Second-Language-Acquisition} & 1\texttimes~\cite{fEDM20variational} \\

\popular 135 & \href{https://www.kaggle.com/datasets/uppulurimadhuri/semeval2013-task-7-sag}{SemEval-2013 (SciEntsBank and Beetle)} -- questions, answers, and labels \cite{SemEval-2013} & 4\texttimes~\cite{fAIED20fooling, sAIED20investigating, fAIED22towards-generating, fAIED23exploration} \\

136 & \href{https://github.com/CAHLR/Serendipitous-Course-Recommendation}{Serendipitous Course Recommendation} -- relationships between courses \cite{fLAK20designing} & 1\texttimes~\cite{fLAK20designing} \\

137 & \href{https://github.com/SebOchs/SAF}{Short Answer Feedback (SAF) Generation} -- short answer grading baseline \cite{Short-Answer-Feedback} & 1\texttimes~\cite{fAIED22towards-generating} \\

138 & \href{https://github.com/converseg/irt_data_repo/tree/master/sim200}{Sim200} -- synthetic array of items and skills for knowledge tracing \cite{sAIED21incorporating} & 1\texttimes~\cite{sAIED21incorporating} \\

139 & \href{https://drive.google.com/file/d/1fMpOPCbGaKBOyVkcoCRCPUbr41AVnPSl/view}{SimPairing} -- student transaction data from studies on group formation \cite{fEDM21exploring} & 1\texttimes~\cite{fEDM21exploring} \\

\popular 140 & \href{https://github.com/chrispiech/DeepKnowledgeTracing/tree/master/data/synthetic}{Simulated-5 (Syn-5)} -- synthetic students’ answering trajectories in tasks \cite{Simulated-5} & 4\texttimes~\cite{sAIED20learning, fAIED21deep, sAIED21incorporating, sAIED21pakt} \\

141 & \href{https://github.com/lan-j/unfair_dataset_generation}{Simulated Bias} -- synthetic features to intentionally induce bias in datasets \cite{fLAK24synthetic} & 1\texttimes~\cite{fLAK24synthetic} \\

142 & \href{https://github.com/itec-hust/singKT-dataset}{singKT} -- music performance assessment data for knowledge tracing \cite{sEDM24singpad} & 1\texttimes~\cite{sEDM24singpad} \\

143 & \href{https://github.com/rezatavakoli/ICALT2020_metadata}{SkillsCommons Resources} -- metadata of educational resources \cite{SkillsCommons-resources} & 1\texttimes~\cite{sLAK21metadata} \\

\popular 144 & \href{https://rajpurkar.github.io/SQuAD-explorer/}{SQuAD: Stanford Question Answering Dataset} -- reading comprehension \cite{SQuAD} & 3\texttimes~\cite{fAIED21automatic, fAIED22towards-human-like, fAIED23scalable} \\

145 & \href{https://pslcdatashop.web.cmu.edu/DatasetInfo?datasetId=308}{Stats2009 (OLI Statistics)} -- statistical reasoning and practice for students \cite{datashop} & 1\texttimes~\cite{sEDM21toward} \\

146 & \href{https://pslcdatashop.web.cmu.edu/DatasetInfo?datasetId=5513}{StatsCloze} -- students' fill-in-the-blank responses to statistical questions \cite{datashop} & 2\texttimes~\cite{fEDM23automated, sEDM23the} \\

147 & \href{https://pslcdatashop.web.cmu.edu/DatasetInfo?datasetId=256}{Stoichiometry} -- student interactions with a tutoring system \cite{datashop, Stoichiometry2} & 1\texttimes~\cite{fAIED24beyond} \\

148 & \href{https://github.com/aied2021TRMRC/AIED_2021_TRMRC_code/tree/main/data/sed}{Student Essay Dataset (SED)} -- essays written by real-world students \cite{fAIED21automatic} & 1\texttimes~\cite{fAIED21automatic} \\

149 & \href{https://www.kaggle.com/datasets/spscientist/students-performance-in-exams}{Students Performance in Exams} -- marks secured by synthetic students \cite{Students-Performance-in-Exams} & 1\texttimes~\cite{fLAK24scaling} \\

150 & \href{https://github.com/machine-teaching-group/edm2024-llm-student-modeling}{StudentSyn} -- synthetic dataset based on solutions to code puzzles \cite{sEDM22from, sEDM24large} & 2\texttimes~\cite{sEDM22from, sEDM24large} \\

\pagebreak

151 & \href{https://github.com/SumnerLab/TalkMoves}{TalkMoves} -- lesson transcripts annotated with discursive moves \cite{TalkMoves} & 1\texttimes~\cite{fAIED23teacher} \\

152 & \href{https://github.com/tbs17/TAPT-BERT/tree/master}{Task-Adaptive Pre-Trained (TAPT) BERT} -- math knowledge components \cite{fAIED21classifying} & 1\texttimes~\cite{fAIED21classifying} \\

153 & \href{https://docs.google.com/forms/d/e/1FAIpQLSfu9IkTLWw97cy8YPrKlhHijBOEYuKjJNiTpVX1UgASbha1AQ/viewform}{Teacher-Student Chatroom Corpus (TSCC)} -- dialogues in language learning \cite{Teacher-StudentChatroomCorpus} & 1\texttimes~\cite{sEDM22the} \\

154 & \href{https://rptsvr1.tea.texas.gov/perfreport/aeis/2008/DownloadData.html}{Texas Academic Excellence Indicator System (AEIS)} -- test results \cite{TexasAEIS} & 2\texttimes~\cite{sEDM24power, sEDM24using} \\

155 & \href{https://allenai.org/data/tqa}{Textbook Question Answering (TQA)} -- questions for 1,076 lessons \cite{Textbook-Question-Answering} & 1\texttimes~\cite{fAIED24real-world} \\

156 & \href{https://pslcdatashop.web.cmu.edu/DatasetInfo?datasetId=5371}{Think-Aloud Chemistry Tutors} -- log data and think-aloud transcripts \cite{datashop, fLAK24using} & 2\texttimes~\cite{fLAK24using, fEDM24using} \\

157 & \href{https://pslcdatashop.web.cmu.edu/DatasetInfo?datasetId=5820}{Think-Aloud Logic Tutor} -- logs of students interacting with an ITS \cite{datashop, fEDM24using} & 1\texttimes~\cite{fEDM24using} \\

158 & \href{https://github.com/iamyuanchung/TOEFL-QA}{TOEFL-QA (Test of English as a Foreign Language)} -- question answers \cite{TOEFL-QA1, TOEFL-QA2} & 1\texttimes~\cite{sAIED24aspect-based} \\

159 & \href{https://github.com/doheejin/aSTS-EI}{TOEFL-QA: aSTS-EI (Semantic Textual Similarity)} -- item similarity \cite{sAIED24aspect-based} & 1\texttimes~\cite{sAIED24aspect-based} \\

160 & \href{https://github.com/umairzahmed/tracer}{TRACER: Targeted RepAir of Compilation ERrors} -- faulty C programs \cite{TracerC} & 1\texttimes~\cite{fAIED20macer} \\

161 & \href{https://www.openml.org/search?type=data&sort=runs&id=42352&status=active}{UCI Student Performance} -- grades and data from Portuguese schools \cite{UCIStudentPerformance} & 2\texttimes~\cite{sEDM21fair, fEDM23investigating} \\

162 & \href{https://github.com/dscodepad/mooc-popularity}{Udemy MOOC Descriptions} -- properties of courses offered by Udemy \cite{sAIED22popularity} & 1\texttimes~\cite{sAIED22popularity} \\

163 & \href{https://pslcdatashop.web.cmu.edu/DatasetInfo?datasetId=126}{USNA Physics} -- logs of student homework done in an ITS \cite{datashop, USNA-Physics2} & 1\texttimes~\cite{fAIED24beyond} \\

164 & \href{https://github.com/psunlpgroup/VerAs}{VerAs: Verify then Assess} -- reports for formative assessment with rubrics \cite{fAIED24veras} & 1\texttimes~\cite{fAIED24veras} \\

165 & \href{https://github.com/sahanbull/VLE-Dataset}{VideoLectures.Net (VLE)} -- metrics for video presentations \cite{fEDM20predicting, sEDM22can} & 2\texttimes~\cite{fEDM20predicting, sEDM22can} \\

166 & \href{https://github.com/vitalsource/data/tree/main/edm-2024}{VitalSource} -- student ratings of automatically generated questions \cite{fEDM24investigating} & 1\texttimes~\cite{fEDM24investigating} \\

167 & \href{https://zenodo.org/records/10252341}{Weights Task: Collaborative Problem Solving} -- transcripts \cite{WeightsTask_CPS} & 1\texttimes~\cite{fEDM24propositional} \\

168 & \href{https://github.com/langcog/wordbankr}{WordBank: Vocabulary Development} -- children's knowledge of words \cite{WordBank-Vocabulary-Development} & 1\texttimes~\cite{fEDM20variational} \\

169 & \href{https://www.kaggle.com/datasets/aljarah/xAPI-Edu-Data}{xAPI Students' Academic Performance} -- student responses to topics \cite{xAPI-Students-Performance} & 2\texttimes~\cite{fLAK24scaling, fEDM23investigating} \\

170 & \href{http://moocdata.cn/data/user-activity#User%20Activity}{XuetangX MOOC Activity of Users} -- students' learning activities \cite{XuetangX-MOOC-activity} & 1\texttimes~\cite{fAIED22balancing-fined-tuned} \\

171 & \href{http://moocdata.cn/data/MOOCCube}{XuetangX MOOCCube} -- concepts, courses, and student behavior \cite{XuetangX-MOOCCube} & 2\texttimes~\cite{fLAK24finding, sEDM21recommending} \\

172 & \href{https://github.com/rezatavakoli/LAK21_metadata}{YouTube Resources} -- metadata of educational videos \cite{sLAK21metadata} & 1\texttimes~\cite{sLAK21metadata} \\

\end{longtable}
}

\section{List of Datasets}
\label{sec:appendix-datasets}

\Cref{tab:datasets} lists all the datasets identified by our survey.

\section{Checklist of Best PRACTICE for Dataset Sharing for Authors}
\label{sec:appendix-checklist}

\Cref{tab:PRACTICE} summarizes the guidelines recommended by our survey. We note that the PRACTICE checklist represents aspirational best practices. When legal, ethical, or institutional constraints prevent full compliance, authors are encouraged to document these limitations transparently.

\newcommand{\checklistitem}{{\huge$\square$} }
\newcommand{\fair}[1]{\textcolor{gray}{[#1]}}

\begin{table}[!ht]
\caption{The 8-item PRACTICE guidelines for open data management in LA research. Each item has 2 actionable checklist sub-items. Also, each item includes a note (in gray) regarding which aspect of the FAIR guidelines~\cite{wilkinson2016fair, EU2018fair} it supports.}
\label{tab:PRACTICE}

\renewcommand{\arraystretch}{1.75}

\newcolumntype{G}{>{\color{gray}\small}c}  
    
\begin{tabular}{G>{\bfseries\Large}c l r}
\hline

\multirow[t]{3}{*}{1} & \multirow[t]{3}{*}{P} & \textit{Prepare for Open Data Sharing Before Collecting the Data.} & \fair{Reusable} \\
& & \multicolumn{2}{l}{\checklistitem Create a research consent form that explains to participants the data anonymization and sharing.} \\
& & \multicolumn{2}{l}{\checklistitem Obtain an IRB approval (if needed). In the application, explain data anonymization and sharing.} \\
\hline

\multirow[t]{3}{*}{2} & \multirow[t]{3}{*}{R} & \textit{Refine and Preprocess the Data for Open Use.} & \fair{Interoperable, Reusable} \\
& & \multicolumn{2}{l}{\checklistitem Anonymize and clean the dataset, especially removing participants' sensitive information.} \\
& & \multicolumn{2}{l}{\checklistitem Have an independent human validator to check for issues. If using AI tools, ensure human oversight.} \\
\hline

\multirow[t]{3}{*}{3} & \multirow[t]{3}{*}{A} & \textit{Add Clear Documentation and Metadata.} & \fair{Findable, Accessible, Interoperable} \\
& & \multicolumn{2}{l}{\checklistitem Describe the data properties, especially the variables and their format in a data dictionary.} \\
& & \multicolumn{2}{l}{\checklistitem Describe the context (e.g., how many participants and data points, where the data were collected, how).} \\
\hline

\multirow[t]{3}{*}{4} & \multirow[t]{3}{*}{C} & \textit{Choose a Trusted and Stable Repository for Longevity.} & \fair{Findable, Accessible, Interoperable} \\
& & \multicolumn{2}{l}{\checklistitem Use a well-established storage, such as Zenodo, to publish the data persistently as a DOI-citable release.} \\
& & \multicolumn{2}{l}{\checklistitem Ensure the data files follow a standard format. Place them in a logical folder structure and describe it.} \\
\hline

\multirow[t]{3}{*}{5} & \multirow[t]{3}{*}{T} & \textit{Transparently License the Data for Open Use.} & \fair{Interoperable, Reusable} \\
& & \multicolumn{2}{l}{\checklistitem Add an open license to the dataset using a standard template, such as CC-BY, allowing reuse.} \\
& & \multicolumn{2}{l}{\checklistitem Ask users to report any breach of license terms, including anonymization issues.} \\
\hline

\multirow[t]{3}{*}{6} & \multirow[t]{3}{*}{I} & \textit{Indicate Open Data Availability in Your Paper.} & \fair{Findable} \\
& & \multicolumn{2}{l}{\checklistitem Link the dataset in a prominent section (e.g., abstract, introduction, conclusion).} \\
& & \multicolumn{2}{l}{\checklistitem Display hyperlinks and references in color to call visual attention.} \\
\hline

\multirow[t]{3}{*}{7} & \multirow[t]{3}{*}{C} & \textit{Citing Your Dataset Should be Easy for Users.} & \fair{Reusable} \\
& & \multicolumn{2}{l}{\checklistitem Add a \say{How to cite} section in the repository, providing both a full-text and BibTeX citation record.} \\
& & \multicolumn{2}{l}{\checklistitem List a valid, official email address where the users can reach you to clarify any questions.} \\
\hline

\multirow[t]{3}{*}{8} & \multirow[t]{3}{*}{E} & \textit{Ensure Responsible Usage and Citation of External Datasets.} & \fair{Reusable} \\
& & \multicolumn{2}{l}{\checklistitem Use others' open datasets responsibly, following the license terms and ethical standards of research.} \\
& & \multicolumn{2}{l}{\checklistitem Give credit to data authors by attributing the dataset in a way the authors prefer.} \\
\hline

\end{tabular}

\end{table}

\end{document}